\documentclass[iop]{emulateapj}

\usepackage{graphicx}
\usepackage{amsmath}
\usepackage{amssymb}
\usepackage{natbib}
\usepackage{color}
\usepackage{enumitem}

\usepackage{txfonts}
\begin{document}

\title{Radiation hydrodynamical models of the inner rim in protoplanetary disks}

   \author{M. Flock\altaffilmark{1,2}, S. Fromang\altaffilmark{2}, N. J. Turner\altaffilmark{1}, M. Benisty\altaffilmark{3}}
   \affil{$^1$Jet Propulsion Laboratory, California Institute of Technology, Pasadena, California 91109, USA}
    \affil{$^2$Laboratoire AIM, CEA/DSM-CNRS-Universit\'e Paris 7,
  Irfu/Service d'Astrophysique, CEA-Saclay, 91191 Gif-sur-Yvette,
  France}
    \affil{$^3$Universit\'e Grenoble Alpes, CNRS, IPAG, 38000 Grenoble, France}
     \email{mflock@caltech.edu}

   \date{}

  \begin{abstract}
Many stars host planets orbiting within a few astronomical units (AU).  The occurrence
rate and distributions of masses and orbits vary greatly with the host star's mass.
These close planets' origins are a mystery that motivates investigating protoplanetary disks' central regions.  
A key factor governing the conditions near the star is the silicate
sublimation front, which largely determines where the starlight is
absorbed, and which is often called the inner rim. We present the first radiation  
hydrodynamical modeling of the sublimation front in the disks around
the young intermediate-mass stars called Herbig Ae stars. The models
are axisymmetric, and include starlight heating, silicate grains
sublimating and condensing to equilibrium at the local, time-dependent temperature and density, and accretion stresses
parametrizing the results of MHD magneto-rotational turbulence
models. The results compare well with radiation hydrostatic
solutions, and prove to be dynamically stable. Passing the model disks
into Monte Carlo radiative transfer calculations, we show that the
models satisfy observational constraints on the inner rims's
location. A small optically-thin halo of hot dust naturally
arises between the inner rim and the star. The inner rim has a substantial radial extent,
corresponding to several disk scale heights. While the
front's overall position varies with the stellar luminosity, its
radial extent depends on the mass accretion rate. A pressure maximum
develops near the location of thermal ionization at temperatures about 1000~K. The pressure maximum is capable of
halting solid pebbles' radial drift and concentrating them in a zone
where temperatures are sufficiently high for annealing to form
crystalline silicates. 
\end{abstract}


\shorttitle{Radiation hydrodynamical models of the inner rim.}
\shortauthors{Flock et al.}

\maketitle

\section{Introduction}
 Thousands of the planets discovered with Kepler and ground-based telescopes orbit
within an astronomical unit of their low-mass host stars \citep{ben14,lis14,joh13}.  
In contrast, intermediate-mass stars around 2~Solar masses more often host super-Jovian 
planets orbiting outside 1~AU \citep{bow10}, while stars of more than 2.7~Solar masses have few or no
  super-Jovians at these distances \citep{ref15}. Our
  understanding of these diverse planetary systems' origins relies on
  our knowledge of the central regions of the disks around all young
  stars.

One key location in the disks is the silicate sublimation front, the
boundary between transparent rock vapor and an opaque dust cloud, and
thus between hot gas lit directly by the star and warm material in the shadow \citep{dul10}.

Another key location, lying near the sublimation front but distinct
from it, is the turbulent front. 
This surface separates material ionized well enough to undergo magneto-rotational instability (MRI)
leading to turbulence, from neutral material that is laminar or
subject to weak turbulence driven by hydrodynamical instabilities \citep{bal98,tur14}. 
Disk annuli experiencing magneto-rotational turbulence have lower surface densities than their
less-turbulent neighbors if the overall flow is in steady state, since
the MRI-active regions' stronger accretion stresses drive the gas
through faster. Thus the turbulent front leads to a nearby local
maximum in the radial profile of the midplane gas pressure \citep{var06,dzy10}.

The pressure maximum collects solid particles, because gas outside the peak orbits slower
than Keplerian, giving the particles a headwind so they drift inward, while gas inside the peak orbits faster than Keplerian, yielding a
tailwind that raises particles' orbits \citep{wei77,hag03,lyr08,lyr09}. 
In particular, the turbulent front can affect the
  distribution of planet-forming solids in intermediate-mass Herbig
  stars' disks \citep{kre09}. The turbulent front's surface density jump also provides favorable
conditions for the growth of cyclonic vortices, which can further
concentrate solid particles along the azimuthal direction 
\citep{bar95,lyr12,fau14b}. Concentrating the solid material makes grain growth efficient \citep{tes14}, and it could lead to in situ planet formation \citep{cha14}, especially at the inner disk \citep{bol14}.
Furthermore, the pressure maximum can halt young planets' orbital
migration \citep{mas06,mat09,lyr10,kre12,bit14,hu15}. The
turbulent front thus plays several roles in planetary systems'
development.\\

The sublimation and turbulent fronts are worth considering together
when modeling the planet-forming environments near young stars,
because they are mutually coupled. The sublimation front affects the turbulent front by governing the starlight absorption and heating. The turbulent front in turn affects the sublimation front through its control over the radial distribution of material. 

Interferometric observations of Herbig Ae/Be stars at near-infrared
(NIR) wavelengths can resolve the sublimation and turbulent fronts'
locations \citep{dul10,kra15}. However, interpreting the measured surface brightnesses in terms of the disk's density and temperature structure remains a challenge.
The ingredients that must be considered include the transfer of the
starlight and the infrared photons re-radiated by the disk, the dust
particles' sublimation and condensation, and the gas vertical
hydrostatic equilibrium \citep{kam09}. Early radiation hydrostatic models had difficulty fitting Herbig stars' spectral energy distributions (SED), especially underpredicting the flux at NIR wavelengths 
\citep{hil92,mil01,mee01,chi01}. Attempts to solve this problem began with a vertical wall
of dust \citep{nat01,dul01}. The vapor between wall and star was optically-thin, letting the starlight fall directly on the wall. This made the wall extra hot, thus extra tall under hydrostatic equilibrium. The puffed-up wall intercepted extra starlight, yielding a higher NIR flux. The next generation of models included the sublimation temperature's density dependence, leading to a rounded shape for the sublimation front: the wall of dust sloped away from the star above and below the midplane, improving the match to disks inclined at a range of angles with respect to our line of sight \citep{ise05,ise06}.\\

Most works have neglected the absorption of starlight in the hot gas
interior to the sublimation front. However, \citet{muz04} suggested that gas between sublimation front and star can be optically
thick enough to push the sublimation front inward significantly.
Gas interior to the sublimation front in the disks around several Herbig Ae/Be stars
was recently detected by CO ro-vibrational emission \citep{ile14} or Br{$\gamma$} emission \citep{men15}. 
An issue not yet addressed by any model is whether the sublimation
front is dynamically stable. Does the sharp temperature jump
destabilize displacements at the front? In this work we investigate
the structure and stability of the sublimation and turbulent fronts,
using radiation hydrostatic and, for the first time, radiation
hydrodynamical models. The models are axisymmetric and treat the transfer of both starlight and infrared radiation, with the opacity linked to the grains' sublimation and condensation. The accretion stresses and heating come from a prescription for MRI turbulence with a switch at the onset of thermal ionization. All the models have surface density profiles corresponding to steady inflow past the two fronts toward the star. We first consider radiation hydrostatic models, then radiation hydrodynamical models. Finally we post-process the results through Monte Carlo radiative transfer calculations to construct synthetic observations of the disk model, which we compare with constraints from observations.\\  

The structure of the paper is a follows. In Section~\ref{sec:method}
we present the radiation hydrostatic method,
the dust and gas opacity and the dust sublimation
module. In Section~\ref{sec:model_hydrostatic} we present a first principle 2D radiation hydrostatic solution to define the inner rim and to show the temperature and density structure, followed by a steady state model in Section~\ref{sec:model_uniform_massflow}. In Section~\ref{sec:freep} we investigate the effects of the model parameter and finally we present the radiation hydrodynamical models in Section~\ref{sec:2D_sim}. In Section~\ref{sec:obs}, we
compare our results with observational constraints and calculate the SEDs of our models, followed by a discussion
(Section~\ref{sec:disc}) and our conclusions (Section~\ref{sec:conc}).  

%

\section{Method}
\label{sec:method}
The aim of the method can be stated as 
the following: given a star with known properties (mass,
radius, luminosity) and a steady state mass accretion rate $\rm \dot{M}$
onto that star, what are the spatial distributions of matter (both gas
and dust) and temperature?

Here, the three important timescales are: the radiative
timescale $\rm t_{\rm rad}$, over which the temperature reaches equilibrium
(given a dust and gas density field), a dynamical timescale associated
with sound waves propagation $\rm t_{\rm dyn}$ over which the vertical disk
equilibrium is set and a long timescale $\rm t_{\rm visc}$ associated with
angular momentum transport in the disk (and often referred to as the
viscous timescale), over which the disk surface density $\rm \Sigma$
evolve. The method we use in the present paper relies on the large
decoupling between the different timescales: $\rm t_{\rm rad}^{\tau=1} \ll t_{\rm dyn} \ll t_{\rm rad}^{\tau \gg 1} \ll t_{\rm visc}$, 
given the optical thickness $\rm \tau$ of the thermal emission in the disk. 

Because of this ordering, the local volume densities remain near vertical 
hydrostatic balance while the total surface density profile evolves. As we 
are interested in the steady-state surface density, we determine $\rm \Sigma$ by assuming a uniform mass accretion rate $\rm \dot{M}$. From the surface density, we calculate the dust and gas densities jointly with the temperature, using an iterative method.

Below in Section~\ref{sec:hyds} we summarize the basic iterative procedure 
described by \citet{flo13}.  In the present paper, we modify this procedure in two key ways.  
First, because we seek to model the sublimation front, we let the dust-to-gas ratio $\rm f_{D2G}$ vary 
with position, greatly affecting the dust opacities.  Second, the sublimation front lies near the thermal ionization 
front, across which magnetic stresses can vary abruptly.  Over the viscous timescale, such a jump in the stress ought to determine the 
surface density profile.  We describe the procedure for computing $\rm f_{D2G}$ in Section~\ref{sec:dust_sub}, and the procedure for estimating the steady-state surface density profile $\rm \Sigma$ in Section~\ref{sec:surf_dens}. Finally, we outline how we compute the dust and gas opacities in Section~\ref{sec:opacities}.
%
%
%
%
%
%
\subsection{Hydrostatic disk structure} 
\label{sec:hyds}
The iterative method solves the coupled equations that
describe vertical hydrostatic balance at a given spherical radius $\rm r$ in the
disk. In a spherical coordinate system $\rm (r,\theta,\phi)$, these
equations are: 
\begin{eqnarray}
\rm \frac{\partial \mathrm{P}}{\partial \mathrm{r}} &=& - \rho \frac{\partial \Phi}{\partial
\rm  r} + \frac{\rho \rm v^2_\phi}{\rm r} \label{eq:P_R} \\
\rm \frac{1}{r} \frac{\partial \mathrm{P}}{\partial \theta} &=&
\rm \frac{1}{\tan{\theta}}\frac{\rho v^2_\phi}{r} \label{eq:P_T} \, ,
\end{eqnarray}
where $\rm \rho$ is the gas density, $\rm v_{\phi}$ is the gas azimuthal
velocity, the gravitational potential $\rm \Phi = GM_*/r$ with the gravitational constant G, stellar mass $\rm M_*$, and $\rm P$ is the thermal pressure that relates to the
temperature $\rm T$ through the ideal gas equation of state:
\begin{equation}
\rm P= \frac{\rho k_B T}{\mu_g u},
\end{equation}
with the mean molecular weight $\rm \mu_g$, the Boltzmann constant $\rm k_B$ and the atomic mass unit $\rm u$.
The temperature in the disk is set by a balance between stellar
irradiation and radiative cooling. For a given density field, the radiation equilibrium is obtained as the steady state solution to the following coupled set of
equations:
\begin{align}
\label{eq:RAD1}
\begin{split}
\rm \frac{1}{\Gamma-1}\partial_t P &= \rm - \sigma_\mathrm{P} \mathrm{c} (a_R T^4 - E_R) - \nabla \cdot F_*, \\
\rm \partial_t E_R - \nabla \frac{c \lambda}{\sigma_\mathrm{R}} \nabla
E_R &= \rm + \sigma_\mathrm{P} \mathrm{c} (\rm a_R T^4-E_R), 
\end{split}
\end{align}
with the adiabatic index $\rm \Gamma$, the radiation energy E$_\mathrm{R}$, the irradiation flux $\mathrm{F}_*$, the Rosseland and
Planck opacity $\rm \sigma_R$ and $\rm \sigma_P$, the radiation constant $\rm a_R=4 \sigma_b/c$ with the Stefan-Boltzmann constant 
$\rm \sigma_b$, and c the speed of light. The flux limiter 
\begin{equation}
\rm \lambda = \frac{2 + R_\lambda}{6 + 3R_\lambda + R_\lambda^2}
\end{equation}
is taken from \citet[][ Eq.~28 therein]{lev81} with 
\begin{equation}
\rm R_\lambda = \frac{|\nabla E_R |}{\sigma_R E_R}.
\end{equation}
The gas is a mixture of molecular hydrogen and helium with solar abundance \citep{dec78} so that $\rm \mu_g =2.35$ and $\rm \Gamma=1.42$. In the Appendix, Table~\ref{tab:constants} provides a list of all the constants we use.

In this work we consider the frequency integrated irradiation flux. $\mathrm{F}_*$ at radius $\mathrm{r}$:
\begin{equation}
\rm F_*(r) = \left (  \frac{R_*}{r}\right )^2  \sigma_b T_*^4 e^{-\tau_*}, 
\label{eq:IRRAD}
\end{equation}
with $\rm T_*$ and $\rm R_*$ being the surface temperature
and the radius of the star. The radial optical depth of the
irradiation flux is defined at each position $\rm \theta$ by:
\begin{equation}
\rm \tau_*(r)=\int_{R_*}^r \sigma_* dr = \tau_0 + \int_{r_0}^r
\sigma_* dr \, ,
\label{eq:TAU}
\end{equation}
where $\rm r_0$ denotes the computational box inner radius and $\rm \sigma_*$ is the opacity at the stellar temperature (see below). $\rm \tau_0$ is
the inner optical depth located between the surface of the star and
$\rm r_0$ and determines how much of the incoming irradiation is blocked
before entering the computational domain. We assume a pure gas disk located between three stellar radii and
$\rm r=r_0$, which gives $\rm \tau_{0}=\kappa_{gas} \rho_{r_0}
(r_0-3R_*)$. $\rm \sigma_* $ is determined by $\rm \sigma_*= \rho_{dust} \kappa_P(T_*) + \rho_{gas}
\kappa_{gas}$. The dust Planck opacity $\rm \kappa_P$ and the gas opacity are defined in Section~\ref{sec:opacities}. 
The inner gas disk edge is assumed to be located at 3 stellar radii, which is the position of the magnetospheric truncation radius for a Herbig type star \citep{muz04}. We note that this raytracing approach neglects scattered starlight. 

The iterative method is summarized below: first, we set the surface density profile $\rm \Sigma(R)$ at the cylindrical radius $\rm R$, and a temperature field $\rm T(r,\theta)$ which is calculated using the optically thin solution. We then calculate 
$\rm \rho(r,\theta)$ and $\rm v_{\phi}(r,\theta)$ by solving
Eq.~(\ref{eq:P_R}) and (\ref{eq:P_T}). We next get the new temperature
profile from the new radiation equilibrium solution of
Eq.~(\ref{eq:RAD1}). We iterate the last two steps 
until we reach convergence. The reader is referred to Section 3.1 of
\citet{flo13} for more details.

\subsection{Dust sublimation}
\label{sec:dust_sub}

The inner disk's structure depends critically on when and where the dust sublimates.  
Sublimation also complicates obtaining converged solutions with iterative methods \citep{kam09,vin12} for at least two reasons.  
First, some material is heated by radiation from outside as well as from inside its orbit. In particular, 
grains star-ward from the front are lit on their night sides by the infrared radiation from their more distant neighbors.  This 
``backwarming'' leaves them hotter than they would be in optically-thin surroundings and this effect is self-consistently included in our method. Second, the front is geometrically 
very thin, since a small column of grains suffices to shield the material beyond from the starlight, making the transition from 
vapor to condensed quite sharp.  High spatial resolution is needed to resolve the layer where the stellar flux is deposited. In our method, we smooth this transition which allows us to overcome this resolution constraint as the models should be suitable for future fully 3D radiation MHD simulations. 

We follow \citet{pol94} and the fitting model of \citet{ise05}\footnote{We note that a slightly different fit was presented by \citet{vin14} (see equation~(A2) therein). We have tested both expressions and found no significant difference on the rim structure.} 
that applies to situations for which the most refractory
grains are silicates. Then the dust sublimation temperature $\rm T_{ev}$ is set by:
\begin{equation}
\rm T_{ev}=2000K \left ( \frac{\rho}{1 g\, cm^{-3}} \right )^{0.0195}
\, . 
\label{eq:ev}
\end{equation}
$\rm T_{ev}$ is then used to calculate the dust-to-gas ratio $\rm f_{D2G}$,
i.e. the ratio between the dust density and the gas
density. We use:
\begin{equation}
\rm f_{D2G}=\left\{
                \begin{array}{ll}
                  \rm f_{\Delta \tau} \left\{ \frac{1-tanh( \left ( \frac{T-T_{ev}}{100K} \right )^3)}{2} \right\}  \left\{ \frac{1-tanh(1-\tau_*)}{2} \right\} &\rm if\, T>T_{ev}\\
                  \rm f_0 \left\{ \frac{1-tanh(20-\tau_*)}{2} \right\} + f_{\Delta \tau}               &\rm if\, T<T_{ev}\\
                \end{array}
              \right.
\label{eq:d2g}
\end{equation}
with the reference dust-to-gas mass ratio $\rm f_0=0.01$ and 
$\rm f_{\Delta \tau}=0.3/(\rho_{gas} \kappa_P \Delta r)$ 
setting the dust amount to account for an optical depth of $\Delta \tau_{*} = 0.3$. Such a value ensures to resolve the absorption of the irradiation at the rim. For $\rm T>T_{ev}$, Eq.~(\ref{eq:d2g}) is similar to the formula of
\citet{kam09} and controls the amount of dust for temperatures above
the sublimation temperature. Note that, in this regime, we also impose
a minimum value of the dust-to-gas mass ratio: $\rm f_{D2G}^{min} =
10^{-10}$. 
The upper limit, in this regime, is given by the value $\rm f_{\Delta \tau}$ which is reached close to $\rm T_{ev}$ and $\rm \tau_*=1$.
For $\rm T<T_{ev}$, Eq.~(\ref{eq:d2g}) limits the dust amount
until the irradiation is absorbed, which is reached close to $\rm \tau_*=20$. 
Finally, the dust-to-gas mass ratio computed by
$\rm f_{D2G}$ is then used to define the total opacity in each grid
cell for the irradiation and for the thermal emission (see Section
~\ref{sec:opacities}). We note that the opacity gradient across the rim could be even more 
gradual than we model here if the dust consists of components with differing sublimation thresholds \citep{mcc13}.  
In addition, the species likely to occur in protostellar disks cover a significant range in sublimation temperature \citep{pol94}.

In the Appendix, we show the robustness of this function by comparing different dust sublimation
functions for $\rm T>T_{ev}$ in Appendix~\ref{app:dsf}. We
demonstrate the importance of the function for $\rm T<T_{ev}$ in the
Appendix~\ref{app:subfunc} and finally, perform a resolution study in Appendix~\ref{app:res}. 

\subsection{Surface density radial profile}
\label{sec:surf_dens}

The disk surface density radial profile is governed by the transport
of angular momentum and evolves over 
long timescales of several thousands of orbits. Here, our strategy is to
use simple scaling laws that come from our understanding of the angular momentum transport to obtain a crude estimate of the equilibrium $\rm \Sigma(R)$, and use it along with the iterative method described above to compute the disk
structure. 

The angular momentum transport in protoplanetary disks inner regions is most likely dominated by MHD turbulence mediated by the MRI. As a result, the flow
is very complex and time dependent. A simple, yet efficient, way to
describe this complex flow is the $\rm \alpha$--prescription, which mimics the effect of the turbulence with a viscosity $\rm \nu_t$ \citep{sha73}. It is customary to 
write this ``turbulent'' viscosity as:
\begin{equation}
\rm \nu_t = \frac{\alpha c_s^2}{\Omega} \, ,
\label{eq:nu_t}
\end{equation}
with the local sound speed $\rm c_s$ and the disk rotation frequency $\rm \Omega=\sqrt{G M_*/R^3}$. 
Using this large scale model
in steady state and far away from the central star,
$\rm \dot{M}$ and $\rm \nu_t$ are related by:
\begin{equation}
\rm \Sigma(R) = \frac{\dot{M}}{3 \pi \nu_t(R)} \, .
\label{eq:sig_mdot}
\end{equation}
For a given value of $\rm \dot{M}$, $\rm \Sigma$ is thus smaller in the
turbulent regions of the disk (where $\rm \alpha$ is large) than in the
laminar parts (where $\rm \alpha$ is small). In this paper, we use
Eq.~(\ref{eq:sig_mdot}) to estimate $\rm \Sigma(R)$. The viscosity is
determined with Eq.~(\ref{eq:nu_t}) in which we specify $\rm \alpha$ using:
\begin{equation}
\rm \alpha = (\alpha_{in} - \alpha_{out} )  \left [ \frac{1-\tanh{\left(
    \frac{T_{MRI}-T}{\Delta T} \right )}} {2} \right ] + \alpha_{out} \, ,
\label{eq:alpha}
\end{equation}
where $\rm T$ stands here for the midplane temperature of the disk. This
formula ensures that $\rm \alpha$ varies smoothly from $\rm \alpha_{in}$ 
at those locations where the MRI is active ($\rm T>T_{MRI}$) to $\rm
\alpha_{out}$ for $\rm T<T_{MRI}$. 
The threshold temperature $\rm T_{MRI}$ for magneto-rotational turbulence is about 1000~K while the exact value depends on the dust-to-gas ratio, the grain sizes, and the gas density \citep{ume88,des15}. The likely ranges in these parameters allow threshold temperatures between about 800 and 1200~K and we study the effect of these values of $\rm T_{MRI}$ on the results in Section~\ref{sec:Tmri}. 

In the remaining of this paper, we use $\rm T_{MRI}=1000\, K$, $\rm \alpha_{in}=1.9 \times 10^{-2}$, $\rm \alpha_{out}=10^{-3}$ and $\rm \Delta T=25$~K. By varying $\rm \Delta T$ from $10$ to $50$~K, we have checked that the influence of its exact value on the
results is small and does not modify our conclusions. A value of $\rm \alpha_{in}$ slightly above 0.01 is justified by MRI simulations in well ionized media
\citep{fro06,dav10,sim11,flo12}. However, the value of $\rm \alpha_{out}$ is
poorly known and depends on the dominant non-ideal MHD term 
\citep{tur14,les14,sim15} and the strength of possible hydrodynamic drivers of turbulence \citep{nel13,kla14,lyr14}. We study the sensitivity of the results to the precise value of $\rm \alpha_{out}$ in Section~\ref{sec:alphaout}. 

\subsection{Opacities}
\label{sec:opacities}
In this paper, we consider gas and dust opacity. We assume $\rm \sigma_R =\sigma_P= \rho_{dust} \kappa_P(T) + \rho_{gas} \kappa_{gas}$ and
we simplify the problem by reducing the frequency-dependent opacities to 3 different frequency-averaged values. The gas
opacity $\rm \kappa_{gas}$, the Planck dust opacity at the rim $\rm \kappa_P(T_{rim})$ and the Planck dust opacity at the stellar temperature $\rm \kappa_P(T_*)$. Such a simplified model compares very well with a more complex model including frequency dependent
irradiation and temperature dependent dust opacity (see Appendix~\ref{ap:freq} for the full comparison). In the following we detail our choices for the dust and gas opacities.

\subsubsection{Dust opacity}

We generate the dust opacity table for different wavelengths using the
MieX code by \citet{wol04}. For details on the dust material, the
opacity calculations and comparison to other dust opacity tables, we
refer to Appendix~\ref{ap:opac}. As mentioned above, two wavelength bands and so two opacity values are important at the rim. The first is the dust opacity at the irradiation temperature. For $\rm T_*=10000~K$ we set $\rm \kappa_P(T_*) = 2100\, cm^2/g$ (exact value $\rm \kappa_P(10000~K) = 2100.3\, cm^2/g$) which is the mean opacity of the dust for the stellar spectrum. We note that we will consider different stellar types and so the value of $\rm \kappa_P(T_*)$ will be adapted
according to the surface temperature of the star. The second frequency averaged opacity
is the dust opacity at the dust sublimation temperature. Typical sublimation temperatures are between 1300~K 
($\rm \kappa_P(1300~K) = 690.1\, cm^2/g$) and 1400~K ($\rm \kappa_P(1400~K) = 717.9\, cm^2/g$). We fix the second dust opacity, which represent the cooling efficiency at the rim, to a value of $\rm \kappa_P(T_{rim}) = 700\, cm^2/g$. 

In this context, the ratio between emission and absorption efficiency $\rm \epsilon$ 
is important as it determines the dust temperature in optically thin, stellar irradiated environment. The small dust particles have a high opacity at short wavelengths
compared to longer wavelengths. They are more efficient in absorbing the shorter wavelength of the stellar radiation than in emitting at longer thermal wavelengths, so they appear hotter than a perfect black body radiator. For our model, the ratio of the emission to absorption efficiency of the dust is then given by $\rm \epsilon = \kappa_P(T_{rim})/\kappa_P(T_*) = 1/3$. Such a value
is typical for a mixture of dust particles with different sizes
\citep{dul10}. Small particles of single size $\rm 0.1 \mu m$, have a value of $\rm \epsilon = 0.08$ \citep{ise05,kam09}, while including larger particles increases the value of $\rm \epsilon$.  

\subsubsection{Gas opacity}

The gas opacity is more difficult to compute as it is dominated by the various line opacities \citep{hel00,dul10}. A fine frequency sampling is required to recover correct values for the mean Planck opacities \citep{mal14}. In addition, such mean opacities can become very high (see Fig. 2 by \citet{mal14}). 
This is because the gas opacity at smaller wavelengths ($\rm \lambda <
0.45 \mu m$) is high due to $\rm H_2$ and CO transitions for
wavelengths shorter than $\rm \sim 0.45 \mu m$ \citep{muz04}.
However, as \citet{muz04} pointed out, there is a lack of absorption in a
wavelength range for $\rm \lambda > 0.45 \mu m$ for the stellar irradiation. Dependent on the molecular abundance and composition, the gas opacity can vary between $\rm 10^{-6} cm^2/g$ and $\rm 1 cm^2/g$ \citep{dul10,mal14} for near infrared wavelengths. In this work, we fix the gas opacity to $\rm \kappa_{gas}=10^{-4} cm^2/g$. The value is chosen so that the radial optical depth $\rm \tau_*$ and vertical optical depth $\rm \tau_Z^{NIR}$ remains small, with  
\begin{equation}
\rm \tau_Z^{NIR}=\int_{-\infty}^{\infty} \sigma_P dz.
\label{eq:TAUZ}
\end{equation}
%
This is very important as otherwise the inner gas disk would block the
irradiation and the rim radius would move too close to the star (see
also Section~\ref{sec:mar}), inconsistent with observations.  
We note also that we use the same gas opacity value for the irradiation
and for the thermal emission, which results in the emission to
absorption efficiency for the gas opacity $\rm \epsilon^{gas} =
1$. Dust will be hotter than gas in optically thin irradiated
regions. In the discussion section we will briefly address again the
effect of the gas opacity on the rim structure and the disk
evolution.\\

\section{The structure of the rim}
\label{sec:model_hydrostatic}

\begin{figure}
\centering
\resizebox{\hsize}{!}{\includegraphics{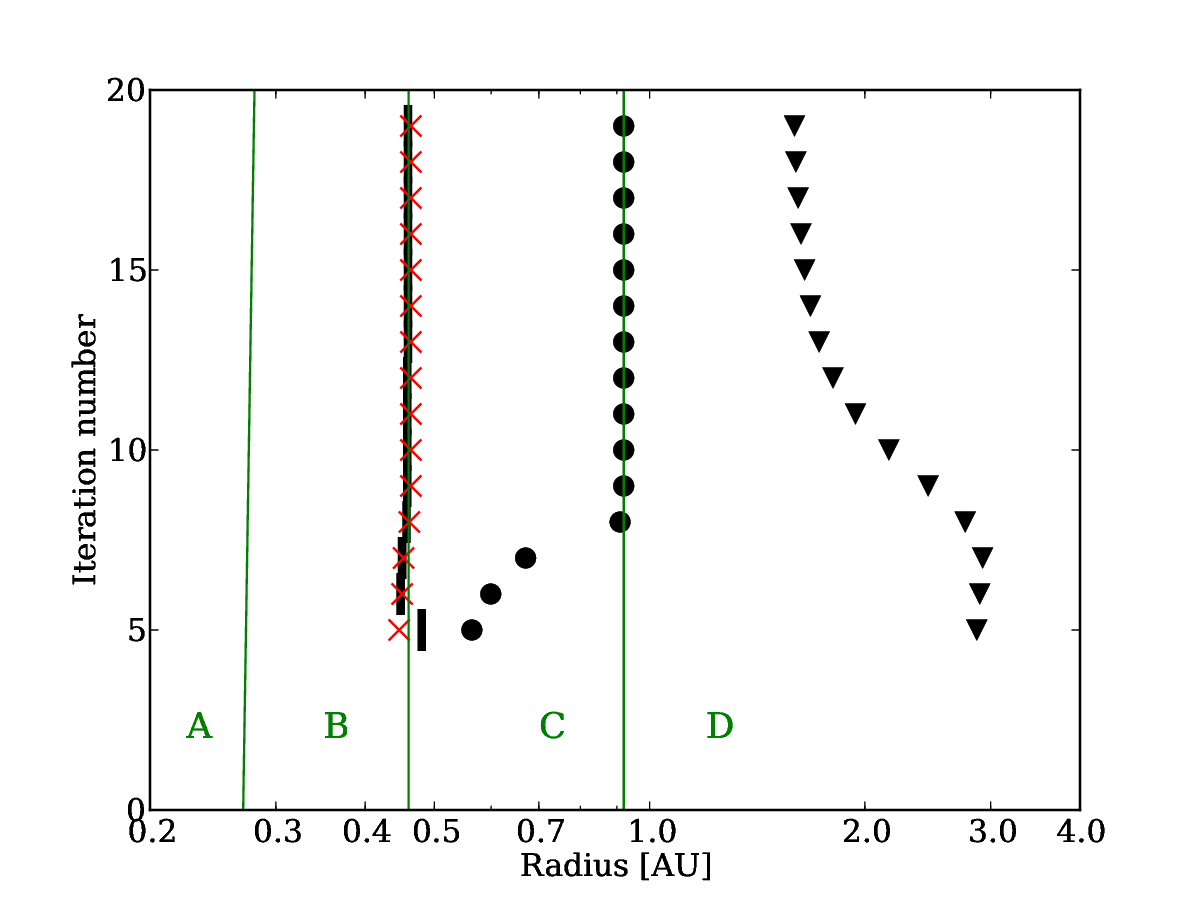}}
\resizebox{\hsize}{!}{\includegraphics{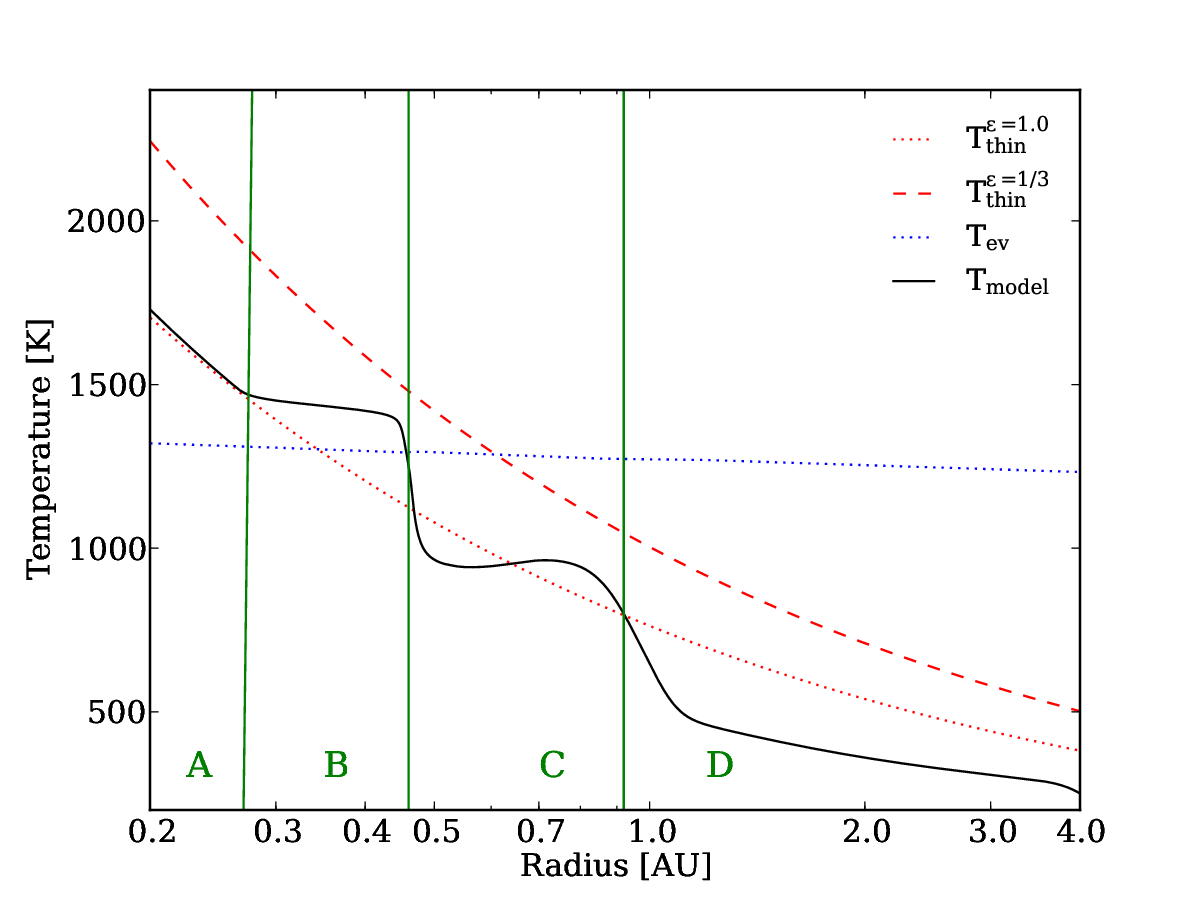}}
\resizebox{\hsize}{!}{\includegraphics{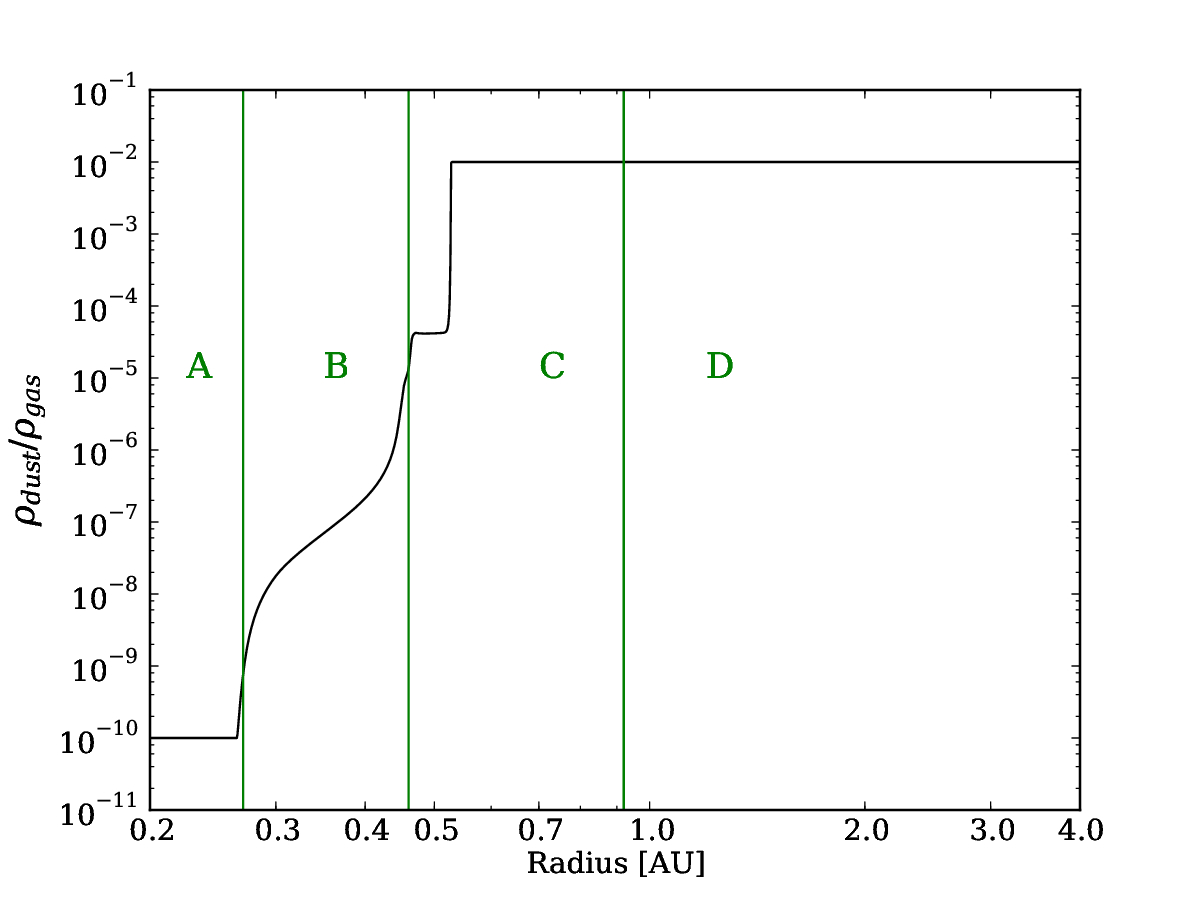}}
\caption{Top panel: Convergence of the \texttt{S100} radiation hydrostatic model.
   The vertical axis is the iteration number.  At each iteration, the 
radius of unit starlight optical depth in the midplane is shown by
a black vertical bar.  Over the first five iterations, the amount 
of dust is increased logarithmically to its final value.  Crosses,
circles and triangles mark where the midplane temperature is 1200, 800 and 400~K, respectively.  
Green vertical lines divide the inner disk into the dust free zone (A), dust halo (B), inner rim (C) and
shadowed region (D).  Middle panel: final midplane radial 
temperature profile (black solid line) in model S100.  Red curves show the temperatures of optically thin gas (dotted) and dust
(dashed).  The blue dotted line shows the dust sublimation temperature.  
Bottom panel: final midplane radial profile of the dust-to-gas mass ratio.} 
\label{fig:temp_thin}
\end{figure}

\begin{figure*}[ht]
\centering
\resizebox{\hsize}{!}{\includegraphics{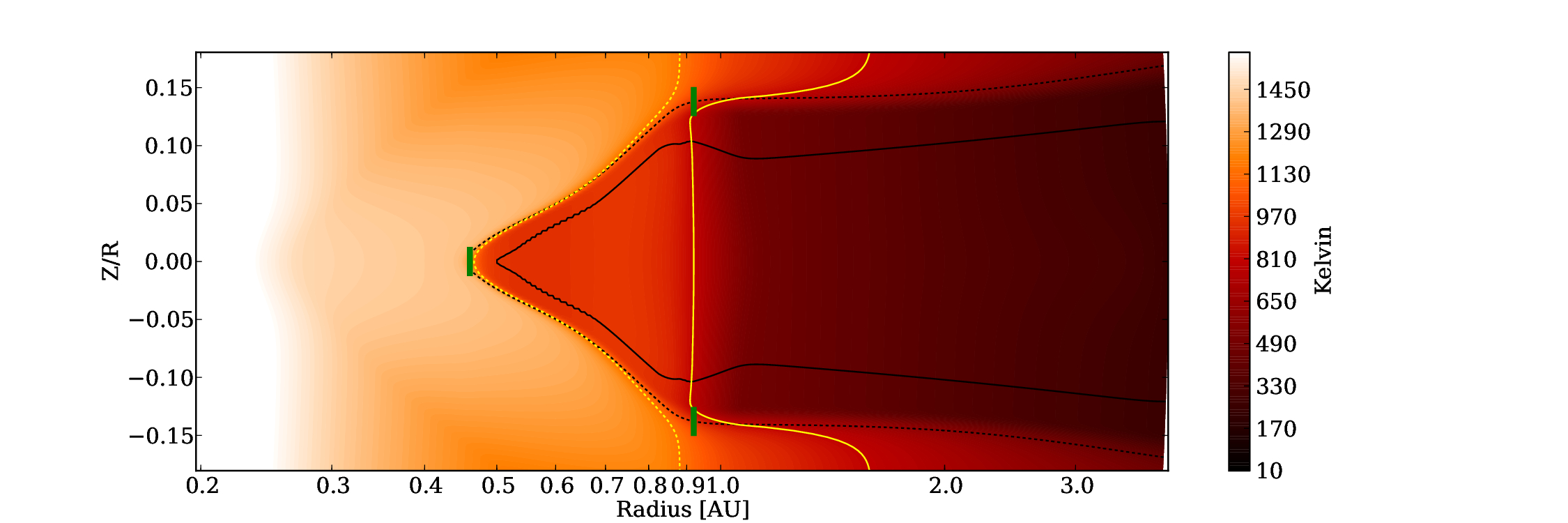}}
\resizebox{\hsize}{!}{\includegraphics{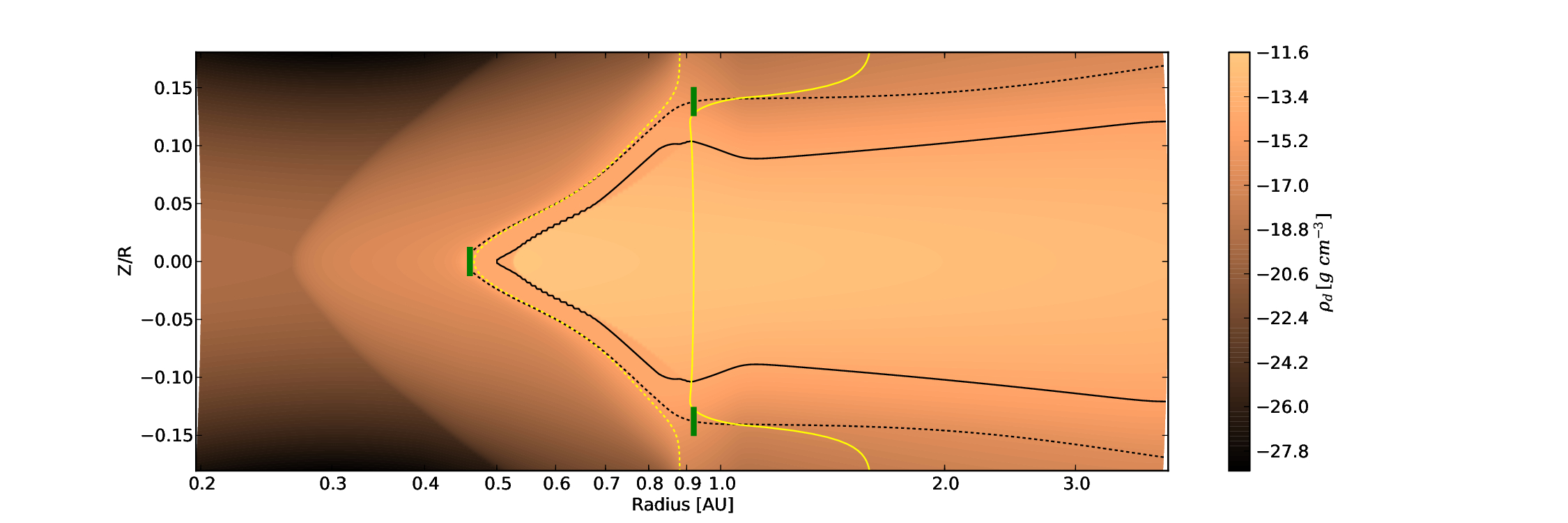}}
\caption{2D profiles of temperature (top) and dust density
  (bottom) in the R$-$Z/R plane, for the converged radiation
  hydrostatic model \texttt{S100}. The black lines indicate the
  optical depth unity for the irradiation (dashed line) and for the
  thermal emission (solid line). The yellow lines show the 1200~K (dashed) and 800~K (solid line)
   temperature contours. The green bars mark the position of $\rm R_{rim}^{in}$ and $\rm R_{rim}^{out}$, which are defined by $\rm \tau_*=1$ at the midplane, and the position in which the incidence angle becomes zero.} 
\label{fig:2dtemp}
\end{figure*} 

In this section we present the typical structure of the rim.
We introduce model \texttt{S100}, which is characterized by a constant surface density of $\rm \Sigma=100\, g\, cm^{-2}$. For this model we neglect the accretion stress and heating. The stellar parameters are the ones of a typical Herbig Ae class star \citep{van98}, with $\rm T_*=10000\, K$, $\rm R_*=2.5\, R_\sun$, and $\rm
M_*=2.5\, M_\sun$. The resulting luminosity is $\rm L_* = 56\, L_\sun$. We use a logarithmically increasing radial grid size, ranging from $\rm 0.2\, AU$ to $\rm 4\, AU$ with a
total of $1280$ grid cells. The vertical extent is $0.36$ radian (20.6$^{\circ}$) and is
composed of $128$ grid cells. The parameters of model \texttt{S100} are
summarized in Table~\ref{tab:info}. \\ 

Fig.~\ref{fig:temp_thin}, top, presents the convergence of model
\texttt{S100}. The plot shows radial optical depth for the irradiation $\rm \tau_*=1$ at the midplane vs. iteration number. The position of three midplane temperatures, 1200~K, 800~K and 400~K, are overplotted. Initially, there is no dust present and
the disk is totally optically thin. In the first five iterations we
increase logarithmically the value of the dust-to-gas mass ratio to its final value. The
results show that the radiation hydrostatic model quickly converges to
a stable rim and temperature structure. 
We note that the optical depth from the inner gas disk in front of our
computational domain (between 0.2~AU and three stellar radii) is $\rm \tau_{0}= 0.14$ at the midplane.

The final converged profiles of the temperature and dust-to-gas mass ratio are
presented in Fig.~\ref{fig:temp_thin} and Fig.~\ref{fig:2dtemp}. 
The radial midplane temperature is plotted in Fig.~\ref{fig:temp_thin}, middle. 
Here, we overplot the dust sublimation temperature from Eq.~(\ref{eq:ev}) and the optically 
thin temperature of the dust and gas. The optically thin
temperature for a given value of absorption to emission efficiency
$\rm \epsilon$ can be calculated with 
\begin{equation}
\rm T_{thin}=\left( \frac{1}{\epsilon} \right )^{0.25} \left( \frac{R_*}{2 r}  \right)^{0.5} T_{*}.
\label{eq:thin}
\end{equation}
Fig.~\ref{fig:temp_thin}, bottom panel, shows the midplane radial profile of the dust-to-gas mass ratio.
The 2D profiles of the temperature and the dust density in the R$-$Z/R%
\footnote{The R$-$Z/R plane has the advantage
that it shows the geometry of the rim along stellar rays
(horizontal lines correspond to the path of individual photons from
the star). The true geometry of the rim in the R$-$Z plane is illustrated in Fig.~\ref{fig:shape} for the more realistic, uniform $\rm \dot{M}$ models.}
plane are plotted in Fig.~\ref{fig:2dtemp}.\\ 

In the converged solution we define four distinct regions, marked with green vertical lines in Fig.~\ref{fig:temp_thin}.
The region A is the dust free disk inward of 0.3~AU. In this zone the
temperature follows the optically thin temperature of the gas ($\rm
\epsilon^{gas}=1$). In zone B, the dust
starts slowly to condense at the rate the temperature drops.  
This region can be also seen as an optically thin dust halo, a region
with a small dust amount. The inner dust halo is clearly visible in Fig.~\ref{fig:2dtemp}, bottom, as well as the rapid increase of the dust density at around $0.5$~AU. We emphasize that this optically thin dust halo appears as a natural outcome of the radiation hydrostatic models.
In this dust halo, the temperature is higher
than the optically thin gas temperature due to the lower $\rm \epsilon$
value of the dust, see Eq.~(\ref{eq:thin}). 

The border between region B and C marks the region in which most of
the dust suddenly condenses, which is the actual beginning of
the rim. This is the position where the irradiation optical
depth passes $\rm \tau_*=1$, see Fig.~\ref{fig:temp_thin}, top. The
final radial position of $\rm \tau_*=1$, between zone B and C, depends
on the global shape of the rim and hence the strength of backwarming, 
and the column of dust and gas in front of the rim. We call this point $\rm R_{rim}^{in}$ from now on.

The rim occupies region C. Here, the optical thickness quickly increases while the
temperature quickly drops. 
To understand the temperature profile across region C, we must 
consider the global shape of the rim, shown in Fig.~\ref{fig:2dtemp}.
The temperature is roughly constant between 0.5 and 0.8~AU. This is due to the high incidence angle at which the starlight strikes the rim surface. 
We define the rim's outer edge as the innermost point where starlight travels parallel to the surface of unit starlight optical depth. This point is marked in Fig.~\ref{fig:temp_thin} by the green vertical line dividing zones C and D, and in Fig.~\ref{fig:2dtemp} by a green bar.  It corresponds to a local maximum in the aspect ratio $\rm Z/R$ of the infrared photosphere $\tau_Z^{NIR}=1$.
%
%
%
%
We call this point $\rm R_{rim}^{out}$ from now on. We note that for model \texttt{S100}, the 1200~K and 800~K contour lines are close to $\rm R_{rim}^{in}$ and $\rm R_{rim}^{out}$. The location of this midplane temperatures compared to the position of $\rm R_{rim}^{in}$ and $\rm R_{rim}^{out}$ changes in the order of 10 \% for the other models presented in this work.

Zone D is the region shadowed by the inner rim, and starts near 1~AU in this model.
Here the temperature drops below the optically thin gas temperature.  We note again, that the shadowing (zero incidence angle) can be seen in Fig.~\ref{fig:2dtemp}, following the
straight irradiation $\rm \tau_*=1$ line between 0.9~AU and 2~AU (green vertical bar). Radially outwards of 2~AU, the disk starts to flare again (non-zero incidence angle). 

In summary, the rim consists of three zones: (1) a hot, optically thin dust halo, 
(2) the starlit rim with its triangular cross-section whose radial extent is several times the density
scale height, and (3) a shadowed, cool zone beyond the rim.  All the models in this work display qualitatively similar structures despite widely differing parameters.  
%
%
%
%
%
%
%
%
\begin{table}
\begin{tabular}{lll}
\hline
Surface density & 100 $\rm g/cm^2$, uniform \\
$\rm N_r \times N_\theta $ & 1280 x 128 \\ 
Cell aspect ratio & $\rm r \Delta \theta / \Delta r \sim 1.2$ \\
$\rm r_{in}-r_{out} : Z/R $ & 0.2-4 AU : $\sim \pm 0.18$\\ 
Stellar parameter & $\rm T_*=10000\, K$, $\rm R_*=2.5\, R_\sun$\\ 
                  & $\rm M_*=2.5\, M_\sun$\\
Opacity & $\rm \kappa_P(T_*)=2100\, cm^2/g$\\
        & $\rm \kappa_P(T_{rim}) = 700\, cm^2/g$\\ 
        & $\rm \kappa_{gas} = 10^{-4}\, cm^2/g$\\
Dust-to-gas mass ratio     & $\rm f_0=0.01$\\
\hline
\end{tabular}
\caption{General setup parameter for the radiation hydrostatic disk model \texttt{S100}.}
\label{tab:info}
\end{table}
%
%
\begin{figure}
\centering
\resizebox{\hsize}{!}{\includegraphics{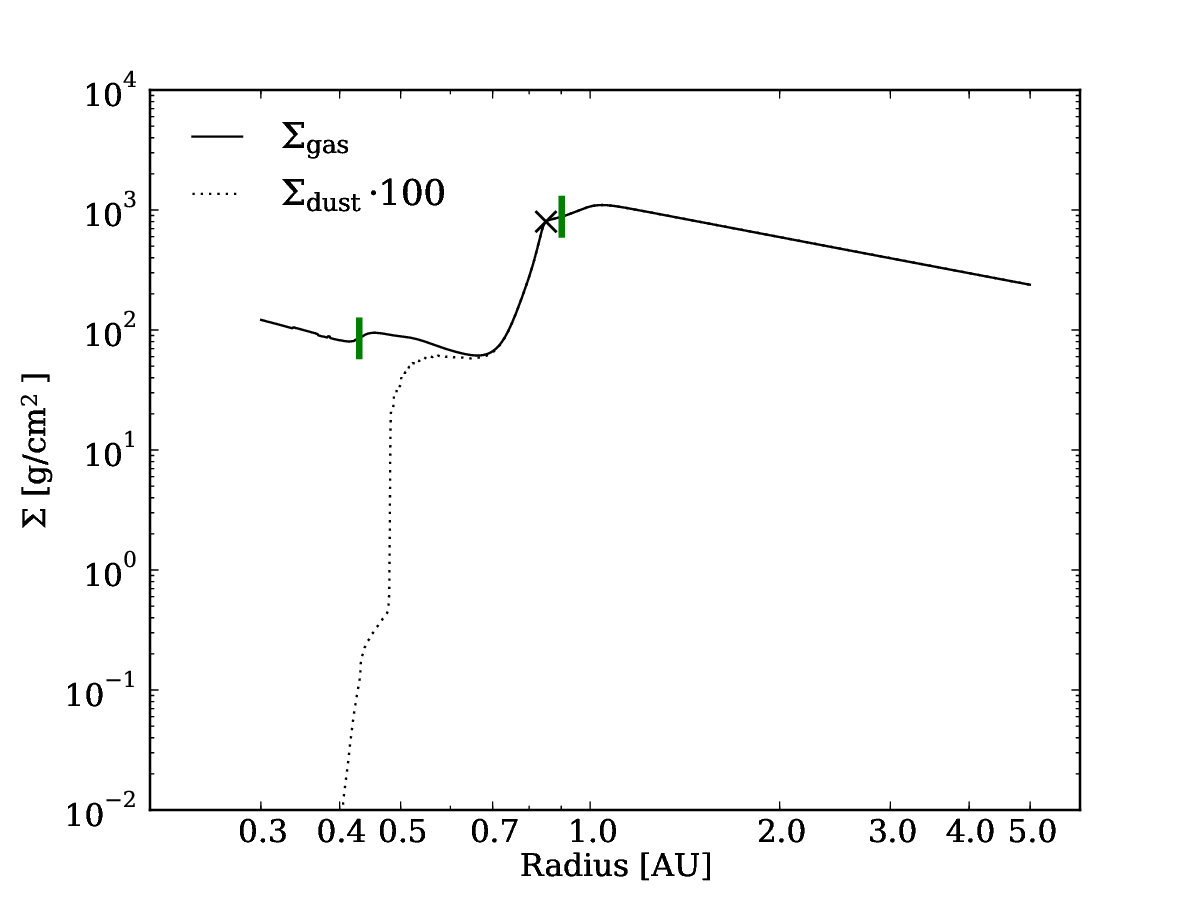}}
\caption{Gas (solid line) and dust (dotted line) surface density
  radial profile for model \texttt{MDe-8}. The green vertical thick bars mark the position of $\rm R_{rim}^{in}$ and $\rm R_{rim}^{out}$ to show the rims extent. The cross marks the location of the pressure maximum.}
\label{fig:sigma_md18}
\end{figure}
\section{Model with a constant $\rm \dot{M}$}
\label{sec:model_uniform_massflow}
We now move to the steady-inflow model with the surface density profile obtained as described in Section~\ref{sec:surf_dens}. In addition to the model described in the previous section, the gas surface density at each iteration is calculated according to Eq.~(\ref{eq:sig_mdot}). We note that in the steady-inflow models, the accretion heating associated with the finite viscosity is neglected.

We first consider model \texttt{MDe-8} with a typical mass accretion rate of $\rm \dot{M}=10^{-8}$ solar mass per year. Previous works have shown that low or even vanishing accretion rates are able to fit the median SED of Herbig type systems \citep{mul12}. The results are presented in Table~\ref{tab:mass_ac} and Fig.~\ref{fig:sigma_md18}. Both gas and dust surface densities show roughly power-law profiles with a jump at $\rm T_{MRI}$. There, the surface densities increase by roughly one
order of magnitude at the location where $\rm \alpha$ drops by roughly one order of magnitude. In the annulus with increasing surface density, the gas is rotating at super-Keplerian velocities. The pressure maximum is located close to $\rm R_{rim}^{out}$ (Fig.~\ref{fig:sigma_md18}, cross symbol). This is the location where we expect solid material to
accumulate. The dust surface density strongly increases at the position of $\rm R_{rim}^{in}$ while otherwise
the dust surface density scales with the gas surface density. Overall, the rim structure looks very similar as presented in Section~\ref{sec:model_hydrostatic}. 
In model \texttt{MDe-8}, the rim's maximum angular extent as seen from the star
is set by the annulus at 0.9 AU, where the starlight-absorbing
surface rises to $\rm Z/R=0.15$, and the near-infrared emitting
surface to $\rm Z/R=0.12$.  These values are similar to radiation
hydrostatic models of the inner rim by \citet{mul12} ($\rm Z/R=0.14$) or by \citet{vin14} ($\rm Z/R=0.11$).

%
\begin{table*}[ht]
\begin{tabular}{lllllll|lllll}
\hline
Model & $\rm \kappa_P^*$ & $\rm \frac{R_*}{R_\sun}$& $\rm \frac{M_*}{M_\sun}$& T$_*$& $\rm \dot{M}$ & $\rm \frac{L_*}{L_\sun}$ & $\rm R_{rim}^{in}$ & $\rm R_{rim}^{out}$& \rm $\rm \tau_Z^{NIR}=1/\tau_r^{*}=1$ & h/r & $\rm P_{max}$\\
\hline
\texttt{S100} & 2100 & 2.5 & 2.5 & 10000 & - & 56 & 0.46  & 0.92  & 0.10/0.14 & 0.04 & -  \\
\hline
\texttt{MDe-8} & 2100 & 2.5 & 2.5 & 10000 & $10^{-8}$ & 56 & 0.42  & 0.89  & 0.12/0.15 & 0.04 & 0.85  \\
\hline
\texttt{MDe-7} & 2100 & 2.5 & 2.5 & 10000 & $10^{-7}$ & 56 & 0.34  & 0.86  & 0.14/0.17 & 0.04 & 0.85  \\
\texttt{MDe-9} & 2100 & 2.5 & 2.5 & 10000 & $10^{-9}$ & 56 & 0.57  & 0.93  & 0.10/0.14 & 0.04 & 0.87 \\
\hline
\texttt{LS21} & 1917 & 2.12 & 2.0 & 8500 & $10^{-8}$ & 21  & 0.13  & 0.48  & 0.10/0.13 & 0.03 & 0.46  \\
\texttt{LS11} & 1784 & 2.0 & 1.8 & 7500 & $10^{-8}$ & 11.3 &  0.09  & 0.31  & 0.09/0.11 & 0.02 &0.28 \\
\hline
\texttt{$\rm \alpha_{out}=\alpha_{in}$} & 2100 & 2.5 & 2.5 & 10000 & $10^{-8}$ & 56 & 0.35  & 0.77  & 0.09/0.13 & 0.04 & - \\
\texttt{$\rm \alpha_{out}=10^{-4}$} & 2100 & 2.5 & 2.5 & 10000 & $10^{-8}$ & 56 & 0.43  & 0.93  & 0.14/0.17 & 0.04 & 0.93 \\
\hline
\texttt{$\rm T_{MRI}=800K$} & 2100 & 2.5 & 2.5 & 10000 & $10^{-8}$ & 56 & 0.41  & 0.97  & 0.12/0.15 & 0.04 & 0.97 \\
\texttt{$\rm T_{MRI}=1200K$} & 2100 & 2.5 & 2.5 & 10000 & $10^{-8}$ & 56 & 0.37  & 0.76  & 0.12/0.15 & 0.04 & 0.43 \\
\hline
\texttt{RHD\_MDe-8} & 2100 & 2.5 & 2.5 & 10000 & $10^{-8}$ & 56 & 0.43  & 0.91  & 0.12/0.15 & 0.04 & 0.90  \\
\texttt{RHD\_MDe-9} & 2100 & 2.5 & 2.5 & 10000 & $10^{-9}$ & 56 & 0.60  & 0.95  & 0.10/0.14 & 0.04 & 0.90 \\
\hline

\end{tabular}
\caption{Input parameter (left columns): model name, irradiation opacity in cm$^2$/g, stellar radius, stellar mass, stellar surface temperature in Kelvin, mass accretion rate in units of solar mass per year, stellar luminosity. Results (right columns): rim position in units of AU, outer rim position in units of AU, maximum height of the NIR/optical surface at $\rm R_{rim}^{out}$ in units of R, scale height $\rm h/r=c_s/v_\phi$ at $\rm R_{rim}^{out}$, location of pressure maximum in units of AU.}
\label{tab:mass_ac}
\end{table*}


\section{Sensitivity to the free parameters}
\label{sec:freep}
There are a number of free parameters that characterize the physical
model. The most important ones are the value of the mass accretion rate and the stellar luminosity which we will investigate in Section~\ref{sec:mar} and Section~\ref{sec:lumi}. The effect of $\rm \alpha_{out}$, the angular momentum transport rate in the dead-zone, and the threshold temperature $\rm T_{MRI}$, are investigated in Section~\ref{sec:alphaout} and Section~\ref{sec:Tmri}, respectively.

\subsection{The influence of $\rm \dot{M}$}
\label{sec:mar}
\begin{figure}
\centering
\resizebox{\hsize}{!}{\includegraphics{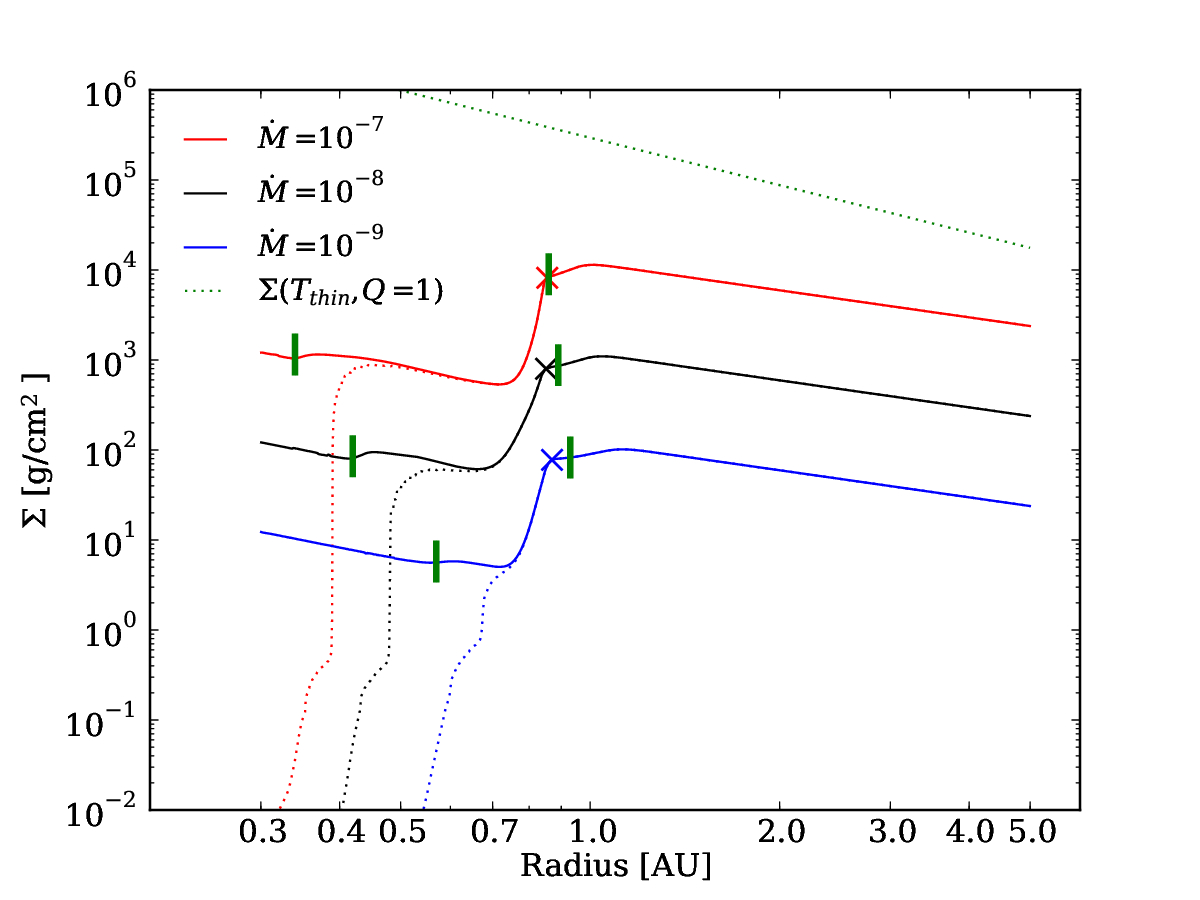}}
\caption{Gas (solid line) and dust (dotted line) surface density
  radial profile for three values of the mass accretion rate
  $\rm \dot{M}$. In each cases, the dust surface density has been
  multiplied by a factor 100. The green vertical thick bars mark the position of $\rm R_{rim}^{in}$ and $\rm R_{rim}^{out}$ to show the rims extent. The crosses mark the location of the pressure maximum. The green dotted line shows the
  critical surface density for which the disk becomes gravitational
  unstable.}
\label{fig:sigma_mdot}
\end{figure}

To study the influence of the mass accretion rates we add two models \texttt{MDe-9} and \texttt{MDe-7} with uniform accretion rates of $\rm \dot{M}=10^{-9}$, and $\rm \dot{M}=10^{-7}$ solar mass per year. 
The results are sumarized in Table~\ref{tab:mass_ac} and Fig.~\ref{fig:sigma_mdot}.
Overall, the  disk structure at the rim looks similar for the three different
mass accretion rates. Both gas and dust surface densities scale with
$\rm \dot{M}$. With increasing mass accretion rate, the position of $\rm R_{rim}^{in}$ moves radially inward, mainly due to the higher optical depth $\tau_0$ and the higher vapor partial pressures in Eq.~(\ref{eq:ev}). Special care should be taken when interpreting the results for the hydrostatic model \texttt{MDe-7}. Here, we had to reduce the value of $\rm \tau_{0}$ by a factor of ten to prevent the inner disk ($\rm r<0.3$~AU) from becoming optically thick, which would move the
rim toward small radii and so out of the computational domain. In addition, for this model the accretion heating is important (see Section~\ref{sec:2D_sim}).

Fig.~\ref{fig:sigma_mdot} includes the upper limit profile of the surface density at which the gravitational instability is triggered. We define the unstable surface density by 
\begin{equation}
\rm \Sigma_{Q=1}=\frac{c_s^{thin} \Omega}{\pi G}, 
\end{equation}
with the Toomre parameter $\rm Q$ and the sound speed at optical thin dust temperature $\rm c_s^{thin}$. As an example, a mass accretion rate of $\rm \dot{M}=10^{-7}\, M_{\sun}\, yr^{-1}$ and an accretion stress such that $\rm \alpha_{out}=10^{-4}$ would render the disk gravitationally unstable
at about 5~AU (see Fig.~\ref{fig:sigma_mdot}).
%
 %
%

\subsection{The influence of the stellar luminosity}
\label{sec:lumi}

It is well known that the stellar luminosity is the most important parameter 
that determines the sublimation front location \citep{dul10}. We
examine its effect on our results by calculating the structure of
disks surrounding stars less luminous than considered above. More
specifically, for the case $\rm \dot{M}=10^{-8}$ solar mass per year, we add the following two models. 
The first uses $\rm L_*=21\, L_\sun$ (for which we have $\rm R_*=2.12\, R_\sun$, $\rm M_*=2.0\,
M_\sun$ and $\rm T_*=8500\, K$) and the second $\rm L_*=11.3\, L_\sun$ (in which
case $\rm R_*=2.0\, R_\sun$, $\rm M_*=1.8\, M_\sun$ and $\rm
T_*=7500\, K$). We named these models \texttt{LS21} and \texttt{LS11},
respectively. Due to the different stellar surface temperature, we have to modify the dust opacities to $\rm \kappa_P(T_*)=1917\, cm^2/g$ and $\rm 1784\, cm^2/g$, and the computational domain radial extent to $\rm 0.1 - 2\, AU$ and $\rm 0.07 - 1.5\, AU$, respectively. All other parameters values are kept fixed. The results are summarized in Table~\ref{tab:mass_ac}.

\begin{figure*}[ht]
\centering
\resizebox{\hsize}{!}{\includegraphics{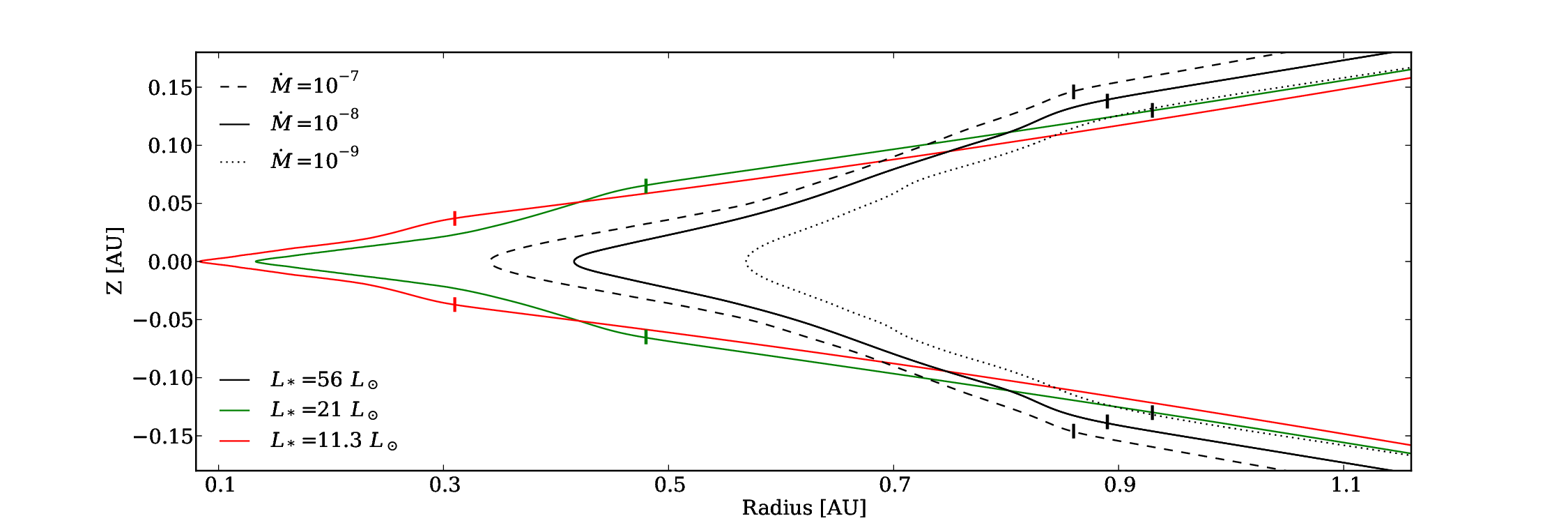}}
\caption{Shape of the irradiation optical depth unity surface for different mass accretion rates and stellar luminosities in the R-Z plane. We conserve the correct aspect ratio to show the true geometry. The vertical thick bars mark the position of $\rm R_{rim}^{out}$.} 
\label{fig:shape}
\end{figure*}

As expected, the rim disk structure clearly moves radially inward when the star luminosity decreases. For example, the location of $\rm R_{\rm rim}^{\rm in}$ moves from $\rm 0.42\, AU$ for a luminosity of $\rm 56\, L_\sun$ to $\rm 0.09\, AU$ for a luminosity of $\rm 11.3\, L_\sun$ in model \texttt{LS11}.    
In addition, the decrease in stellar mass leads to an overall thinner disk at the position of $\rm R_{rim}^{out}$. As this position, the scale height $\rm h/r=c_s/v_\phi$ gradually decreases from $0.04$ for the fiducial case, to $0.033$ and $0.026$.  
Fig.~\ref{fig:shape} summarizes the true geometry of the rims surface in the R-Z plane for the previous models. 

\subsection{The influence of $\rm \alpha_{out}$}
\label{sec:alphaout}
\begin{figure}
\centering
\resizebox{\hsize}{!}{\includegraphics{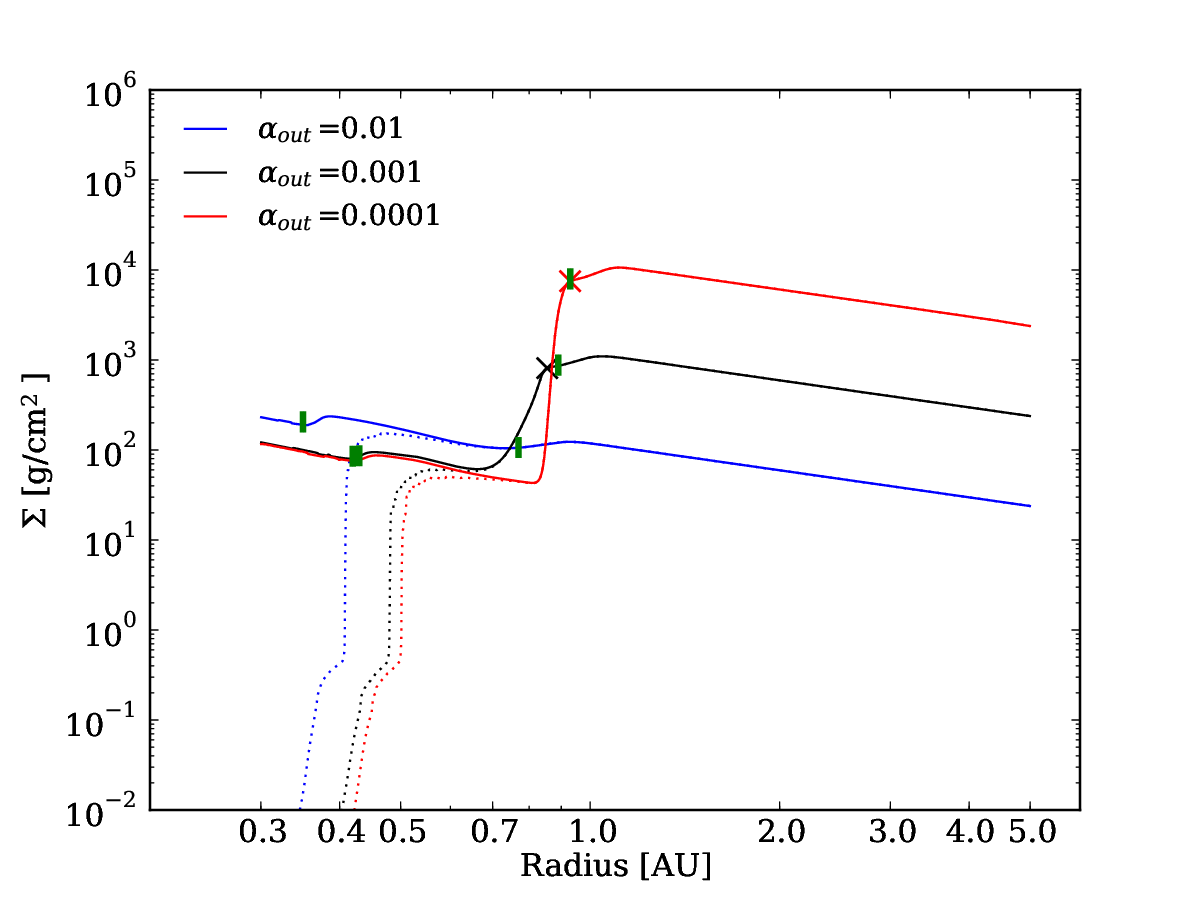}}
\caption{Radial surface density profile of the gas (solid line) and
  dust multiplied by a factor 100 (dotted line), for the models with
  modified $\rm \alpha_{out}$. The vertical thick bars mark the position of $\rm R_{rim}^{in}$ and $\rm R_{rim}^{out}$. The crosses mark the location of the pressure
  maximum.} 
\label{fig:alpha_comp}
\end{figure}
In this section we investigate the effect of the value of $\rm \alpha_{out}$ on the rim structure. To do so, we use 
the parameters of model \texttt{MDe-8} and considered in addition the
two cases given by $\rm \alpha_{out}=\alpha_{in}$ and $\rm
\alpha_{out}=10^{-4}$ (we remind the reader that $\rm
\alpha_{out}=0.001$ in model \texttt{MDe-8}). The first case corresponds to a fully turbulent disk (i.e. without a dead zone) and
the second describes a nearly laminar dead zone. The
results are summarized in  Table~\ref{tab:mass_ac} and
Fig.~\ref{fig:alpha_comp}. The model \texttt{$\rm \alpha_{out}=10^{-4}$} yields
results that are similar to model \texttt{MDe-8}. The rim radius
$\rm R_{\rm rim}^{\rm in}$ moves only slightly outward (from $0.42$ to
$0.43$~AU). Changes remain modest but are more important for the
case $\rm \alpha_{out}=\alpha_{in}$. Here, the most important difference is the absence of a pressure maximum. $\rm R_{\rm
  rim}^{\rm in}$ decreases from $0.42$ to $0.35$~AU because of the reduced backwarming of the dust wall. 

Overall, we conclude that the effect of varying $\rm \alpha_{out}$ remains
modest. This is because the rim is mainly located where $\rm T> 1000 K$ whereas the surface density changes due to varying
$\rm \alpha_{out}$ occur at locations where $\rm T < 1000$~K. 

\subsection{The influence of $\rm T_{MRI}$}
\label{sec:Tmri}
For the last parameter comparison we investigate the effect of the value of $\rm T_{MRI}$ 
on the rim structure. To do so, we use the parameters of model \texttt{MDe-8} and consider in addition the
two cases given by $\rm T_{MRI}=800\, K$ and $\rm T_{MRI}=1200\, K$. The
results are summarized in Table~\ref{tab:mass_ac}. The case $\rm T_{MRI}=800\, K$ is similar to model \texttt{MDe-8}. The outer rim radius
$\rm R_{\rm rim}^{\rm out}$ moves slightly outward (from $0.89$ to $0.97$~AU) due to the surface density increase at larger radii. The same happens with the position of the pressure maximum, which moves outward from $\rm P_{\rm max}=0.85$ to $\rm P_{\rm max}=0.97$~AU. 

The case $\rm T_{MRI}=1200\, K$ has a stronger effect on the rim profile. In this case, the higher surface density at higher temperatures shifts the rim from $\rm R_{\rm rim}^{\rm in}$ at $0.42$~AU to $0.37$~AU and from $\rm R_{\rm rim}^{\rm out}=0.89$~AU to $\rm R_{\rm rim}^{\rm out}=0.76$~AU. The pressure maximum shifts even more, from $0.85$~AU in model \texttt{MDe-8} to $0.43$~AU.
For model $\rm T_{MRI}=1200\, K$, we set the computational domain radial extent to $\rm 0.2 - 3\, AU$. In Section~\ref{sec:sedah}, we compare the previous models with the observational contraints.
\section{Radiation hydrodynamical simulations}
\label{sec:2D_sim}
Finally, we investigate the stability of the rim, using long term 2D radiation hydrodynamical simulations. The hydrostatic models are used as initial conditions. Our goals are twofold: in addition to verifying that the disk remains in
steady state for thousands of dynamical times, we also want to
investigate the effect of accretion heating on the disk
structure. The numerical setup is almost identical to that used by
\citet{flo13} and is briefly recalled in Appendix~\ref{ap:rmhd} along with a
more detailed description of the modifications required for the simulations to be completed.

\begin{figure}
\centering
\resizebox{\hsize}{!}{\includegraphics{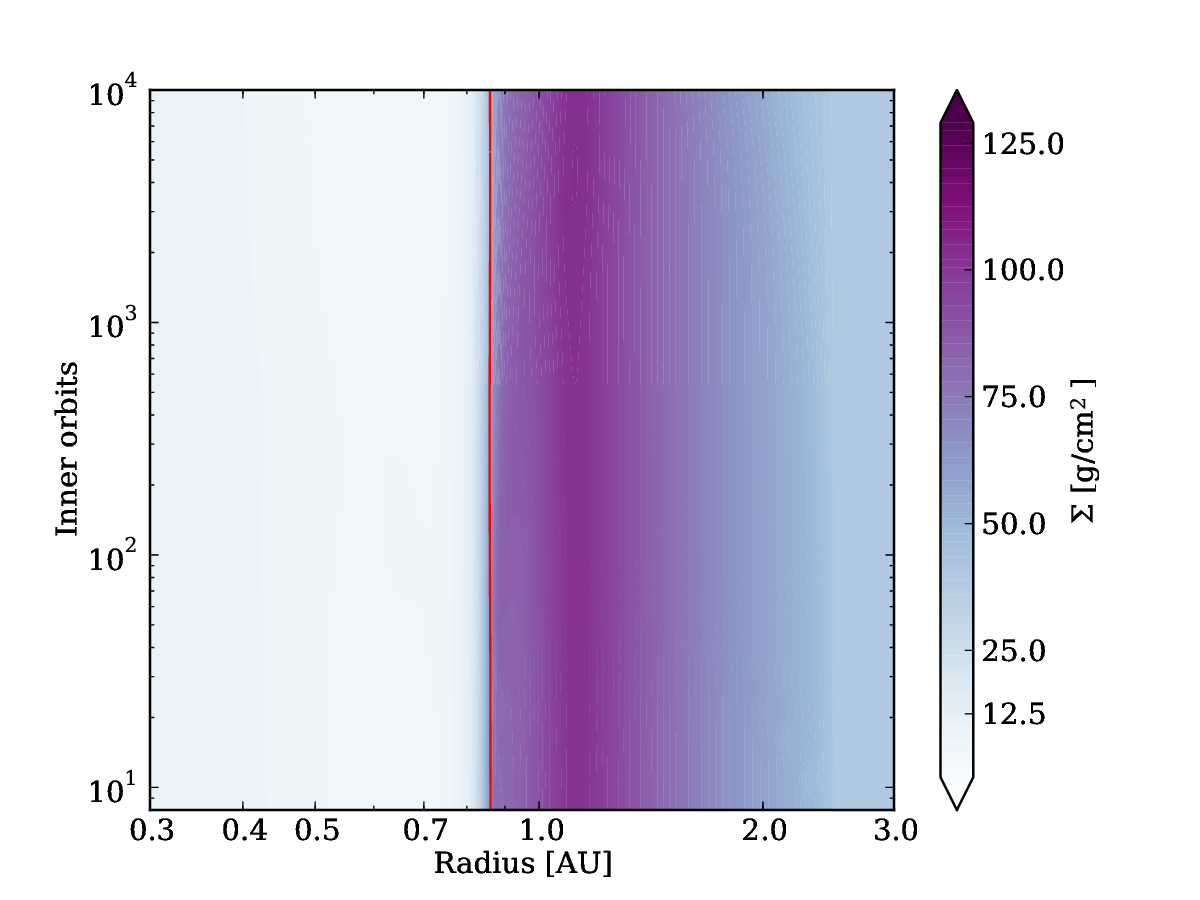}}
\resizebox{\hsize}{!}{\includegraphics{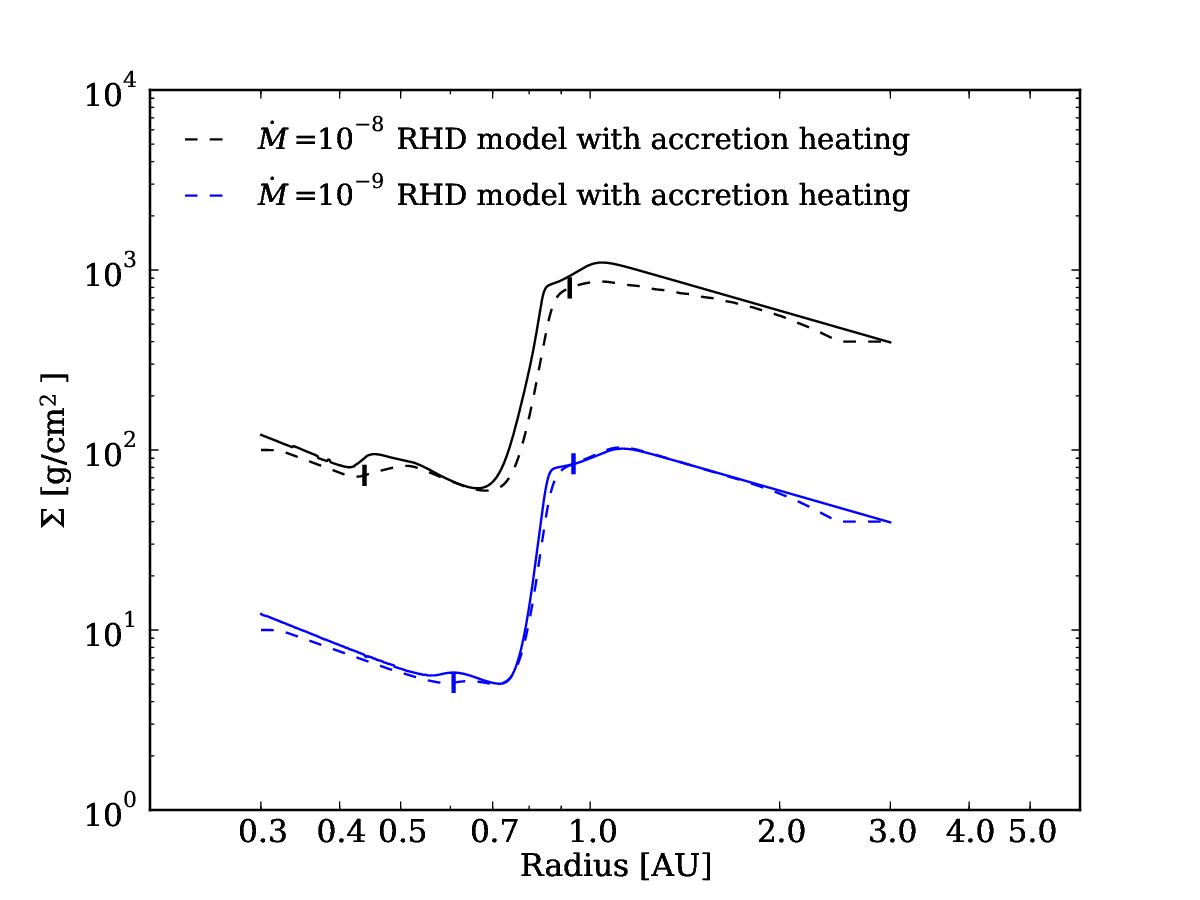}}
\caption{Top: Time evolution of the radial surface density profile for the radiation hydrodynamical model \texttt{RHD\_MD1e-9}. Bottom: Radial surface density profiles of the radiation hydrostatic models \texttt{MD1e-8} and \texttt{MD1e-9}  (solid lines) compared to the final snapshots of the radiation hydrodynamical models \texttt{RHD\_MD1e-8} and \texttt{RHD\_MD1e-9} (dashed lines). The vertical thick bars mark the position of $\rm R_{rim}^{in}$ and $\rm R_{rim}^{out}$.}
\label{fig:sig_acc}
\end{figure}

\begin{figure}
\centering
\resizebox{\hsize}{!}{\includegraphics{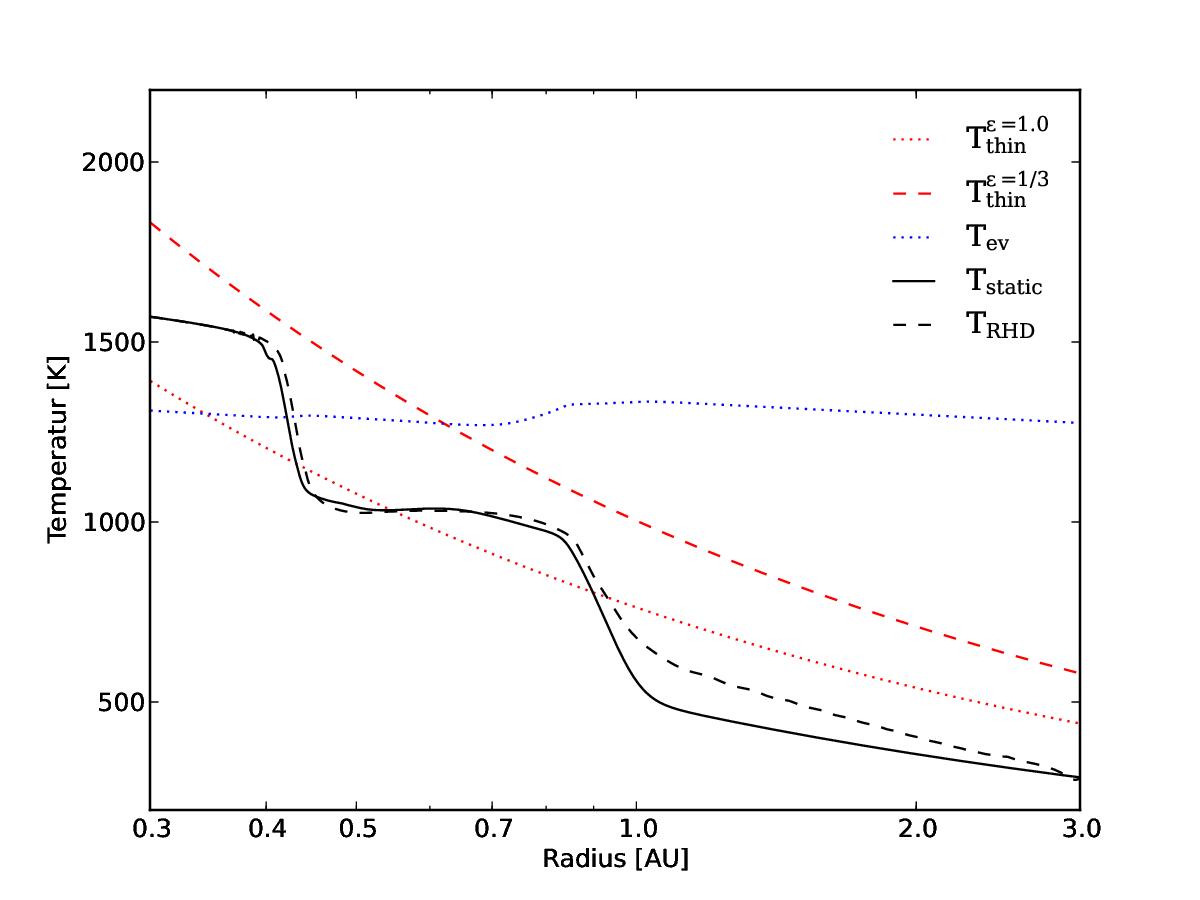}}
\caption{Radial midplane temperature profile of model \texttt{MD1e-8} (black solid line) and model \texttt{RHD\_MD1e-8} (black dashed line). The red lines correspond to the optical thin temperature of the gas (red dotted line) and the dust (red dashed line). The blue line shows the sublimation temperature of the dust.}
\label{fig:temp_evo}
\end{figure}

We chose the models \texttt{MDe-8} and \texttt{MDe-9} to match the mass accretion rate of typical Herbig star models \citep{mul12}. The RHD simulations \texttt{RHD\_MD1e-9} and
\texttt{RHD\_MD1e-8} were integrated for a runtime of $10000$ inner orbits. In both simulations, the final 
disk structure after $10000$ inner orbits is almost identical to the initial state. In both models, the radial velocity fluctuations
remain small, of the order of $\rm 10^{-3} c_s$ in the midplane. For
illustrative purpose, we present in Fig.~\ref{fig:sig_acc} (top
panel) a spacetime diagram showing the evolution of the surface density radial profile 
for model \texttt{RHD\_MD1e-9}. The final surface density
profiles of both models are compared in Fig.~\ref{fig:sig_acc}
(bottom panel) to the initial static disk solution. The results show that the final 
profiles are very close to a passive disk solution (i.e. without
accretion heating), especially for model \texttt{RHD\_MD1e-9}. For
model \texttt{RHD\_MD1e-8}, the surface density adapts to a slightly
different equilibrium as the effect of accretion heating becomes
visible. The radial midplane temperature of model \texttt{RHD\_MD1e-8}
at this final state is plotted in Fig.~\ref{fig:temp_evo},
overplotting also the initial temperature profile. The effect of the
accretion heating becomes visible especially in the shadowed
region. However we note that overall the density and temperature
structure of model \texttt{RHD\_MD1e-8} remains similar to the
radiation hydrostatic model \texttt{MD1e-8}.\\ 

\section{Comparison with observational constraints}
\label{sec:obs}
In this section we compare our results with observational
constraints. We start by presenting the SEDs of our models in Section~\ref{sec:sed}. We investigate the effect of the model parameters on the SED in Section~\ref{sec:sedah}. Finally, we compare our results for the rim radius with observational constraints in Section~\ref{sec:rim_radius} and construct synthetic images in Section~\ref{sec:syni}.

\subsection{Spectral Energy Distribution}
\label{sec:sed}

\begin{figure}
\centering
\resizebox{\hsize}{!}{\includegraphics{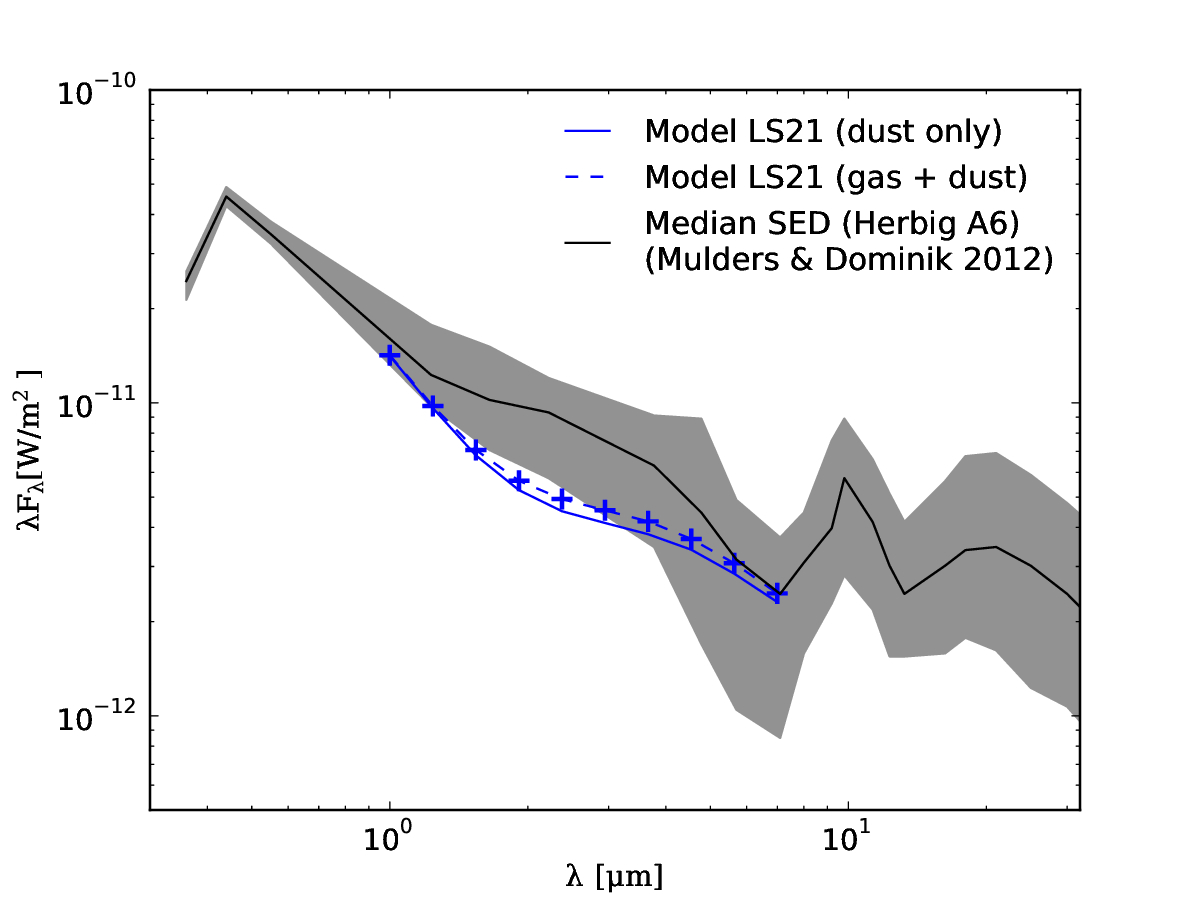}}
\caption{SED of the sample of Herbig stars (median spectral type A6) by
  \citet{mul12} (black lines) with the upper and lower quartile (gray
  shade). The SED of model \texttt{LS21} is shown with the pure dust
  component (solid blue line) and with including an additional gas
  component (dashed line). The blue crosses show the seven individual
  wavelengths calculated for our models.} 
\label{fig:sed}
\end{figure}

We use the Monte Carlo radiative transfer code RADMC3D \citep{dul12} to calculate the SED associated with the disk structures computed above. We consider seven
individual wavelengths between 1 and 7 $\rm \mu m$ to cover the 
regime corresponding to our domain size and temperature range. For the dust opacity we use the wavelength dependent table as shown in Fig.~\ref{fig:Opac}. For the calculation of the SED, we assume a disk inclination of $45^\circ$ and we scale the results to a distance of $122$ parsec (which corresponds to the distance of the Herbig star HD~163296). We compare our results with the Herbig star sample described 
by \citet{mul12} for which model \texttt{LS21} matches the Herbig A6
median stars' luminosity, surface temperature and stellar mass (see Appendix B therein). We calculate the SED using two
strategies. In the first, we consider the contribution of the dust component only. 
In the second, we add the gas component, assuming a gray gas opacity. 
The results of both models are plotted in Fig.~\ref{fig:sed} and compared with the sample from observations. The NIR emission of model \texttt{LS21} is below the observational median, especially at two microns, where the emission is lower by a factor of two. The gas only contributes a few percent of additional flux in this regime. 

\subsection{Effect of accretion heating, $\rm \alpha_{out}$ and $\rm T_{MRI}$ on the SED}
\label{sec:sedah}
In this section we investigate the effects of the accretion heating, $\rm \alpha_{out}$ and $\rm T_{MRI}$ on the SED. 
Overall the effects remain small. We compare the SED of the hydrostatic model \texttt{MD1e-8} with the profile of the radiation hydrodynamical model \texttt{RHD\_MD1e-8}, using the simulation output we obtained at the last timestep. The two resulting SEDs are very close, with differences less than $5\%$. 

Similarly small changes are seen for model \texttt{$\rm \alpha_{out}=10^{-4}$}, which has a higher dust density at the outer rim position.
For the case \texttt{$\rm \alpha_{out}=10^{-4}$} the emission at 4 microns is greater by 8\% than the model \texttt{MD1e-8} with $\rm \alpha_{out}=0.001$. 
For model \texttt{$\rm \alpha_{out}=\alpha_{in}$}, there is no surface density increase at the outer rim and the emission at 4 microns decreases by 17\% compared to the model with $\rm \alpha_{out}=0.001$.
Comparing with the observational constraints, the models would favor a strong drop of $\rm \alpha_{out}$ but the changes remain small and we cannot rule out the possibility of a constant $\rm \alpha_{out}$ from the SED alone.

Finally model \texttt{$\rm T_{MRI}=800K$} shows a 6\% decrease of emission at 4 microns as the surface density jump moves radially outward, affecting the emission at longer wavelengths. Model \texttt{$\rm T_{MRI}=1200K$} shows a small increase of 5\% at 2 microns but a larger decrease of 10\% at 5 microns compared to model \texttt{MD1e-8} with $\rm T_{MRI}=1000K$.

\subsection{Rim radius}
\label{sec:rim_radius}

\begin{figure}
\centering
\resizebox{\hsize}{!}{\includegraphics{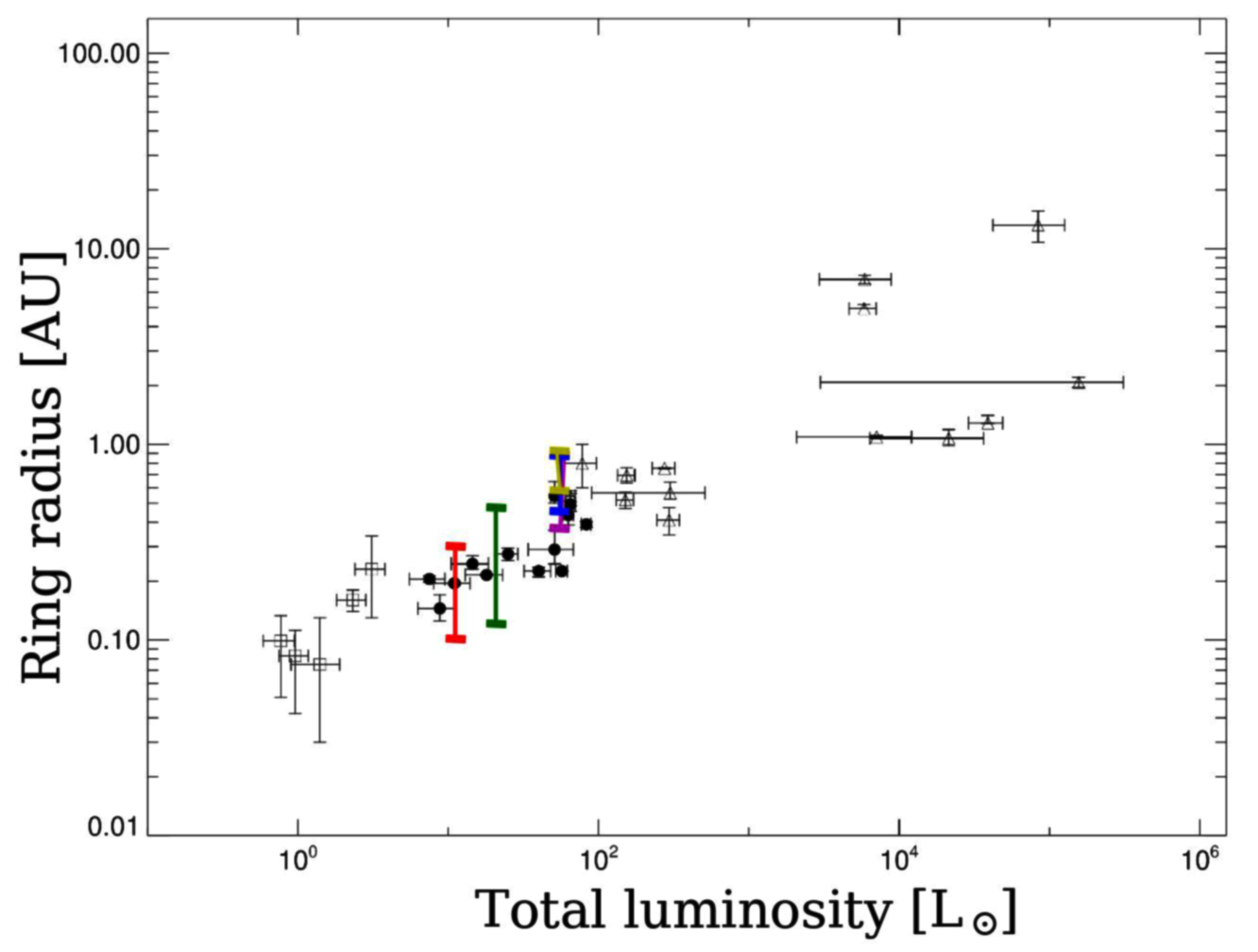}}
\caption{Inner dust ring radius over luminosity relation. The square, circle and rectangle symbols represent T Tauri, Herbig Ae and Herbig Be stars, respectively, adapted from \citet{dul10} and \citet{mil07}. We overplot our results by
  colored vertical lines, showing the radial extent $\rm  R_{rim}^{in}$ and $\rm R_{rim}^{out}$ in
  Table~\ref{tab:mass_ac}. The different models are, \texttt{LS11}
  (red), \texttt{LS21} (green), \texttt{MDe-9} (yellow),
  \texttt{MDe-8} (blue) and \texttt{MDe-7} (magenta). We note that for the \texttt{MDe-X} models, the value of $\rm  R_{rim}^{out}$ is similar.} 
\label{fig:ringrad}
\end{figure}

As we have seen, the sublimation front takes a pointed shape
spread over several pressure scale heights in radius. It is thus not
straightforward to define the exact radial position of the rim. In this section, we compare the observational determined radii presented by \citet{mil07} and \citet{dul10} with the rim radial extent $\rm R_{rim}^{in}$ and $\rm R_{rim}^{out}$ from our models (see Table~\ref{tab:mass_ac}).
As shown on Fig.~\ref{fig:ringrad}, the theoretical and observational
rim radii compare very well for the range of stars
luminosity we explored. This comparison highlights once more that the
actual rim becomes more radially extended with higher mass accretion rate, as shown by the vertical length of the bars in Fig.~\ref{fig:ringrad} for the \texttt{MDe-X} models. By contrast, the actual ring position is determined by the central star luminosity and agree with the systematic trend seen in the observations. We note that we did not include the effect of accretion luminosity as this remains small for the considered systems \citep{mul12}.

Overall, the SED and the position of the rim from our models are in broad agreement with previously published models by \citet{dul10,mul12} which do not include the gas between the rim and the star. As previous models have shown, there is still a lack emission at two micron wavelength by roughly a factor of two. Neither the presence of a small dust halo, the contribution of the gas, accretion heating nor changing the surface density or the position of the surface density increase are able to solve that problem. As recently shown by \citet{tur14}, the effect of magnetic pressure, namely a thickening of the disk at the rim location, remains a possible solution.

\subsection{Synthetic images}
\label{sec:syni}
\begin{figure}
\resizebox{\hsize}{!}{\includegraphics{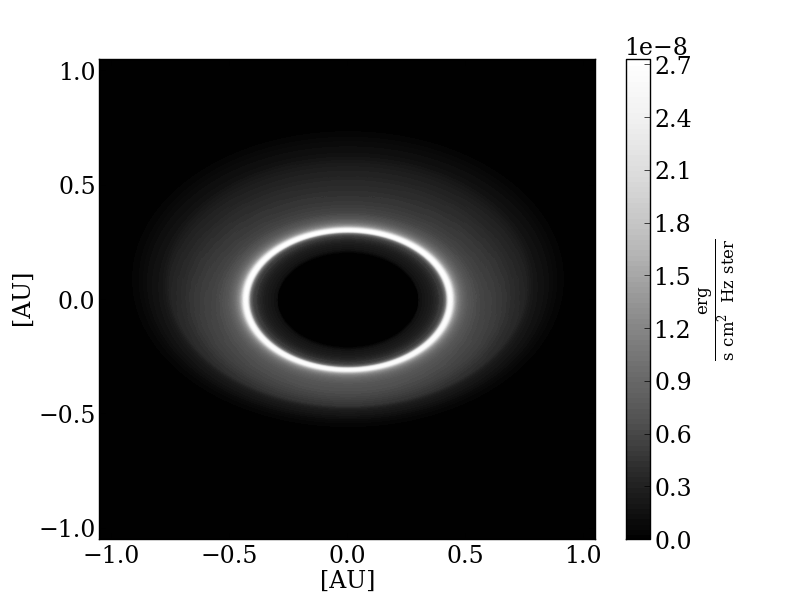}}
\resizebox{\hsize}{!}{\includegraphics{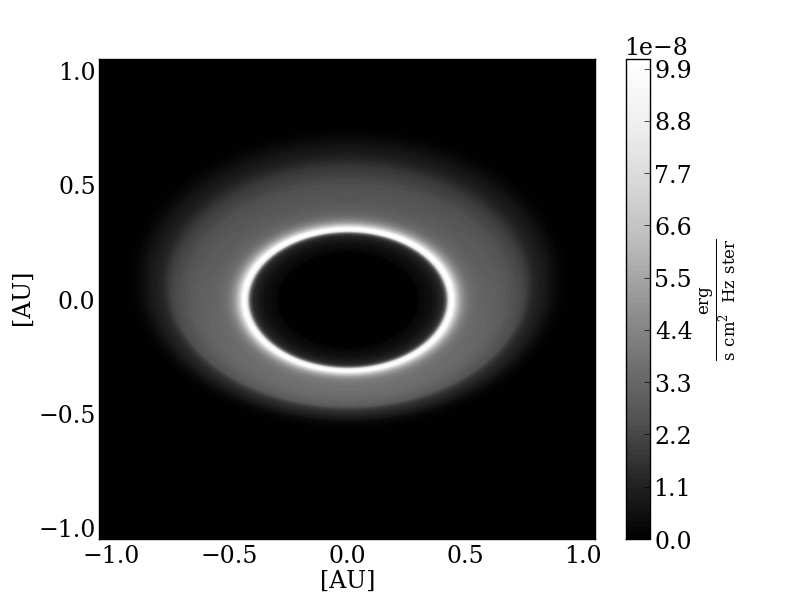}}
\resizebox{\hsize}{!}{\includegraphics{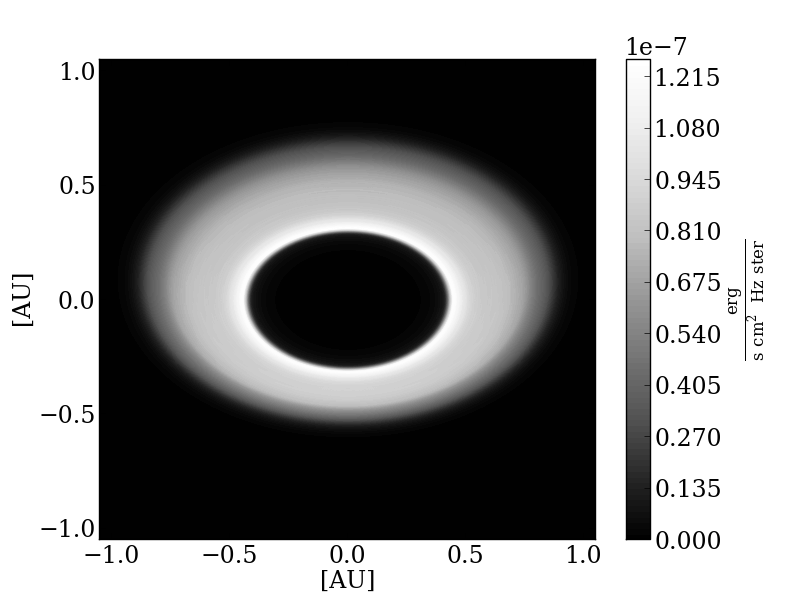}}
\caption{ Synthetic images of the final output from the
    radiation hydrodynamical model \texttt{RHD\_MD1e-8}, viewed 45$^\circ$ from
    face-on. The intensity maps correspond to 1.25 (top), 2.2 (middle) and 4.8 $\rm \mu$m (bottom).} 
\label{fig:sed_im}
\end{figure}

\begin{figure}
\resizebox{\hsize}{!}{\includegraphics{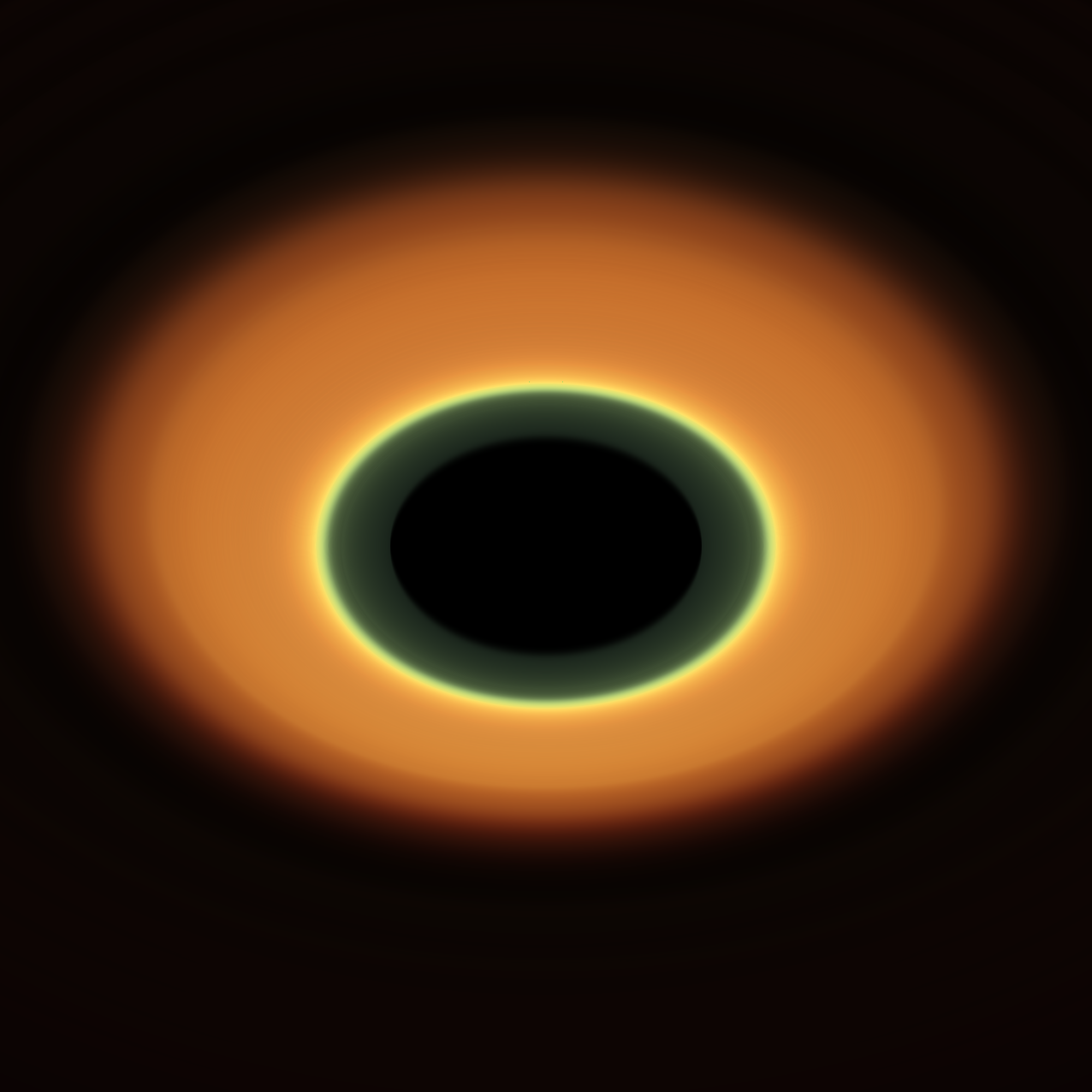}}
\caption{ Synthetic color image of the final output from the
    radiation hydrodynamical model \texttt{RHD\_MD1e-8}, viewed 45$^\circ$ from
    face-on. The blue, green and red
    channels come from the 1.25, 2.2 and 4.8-$\mu$m bands
    respectively.  The three channels share a common, linear intensity
    scale determined by the extrema of the 4.8-$\mu$m image.} 
\label{fig:sed_im_2}
\end{figure}

To finally provide a realistic view of the rim, we constructed synthetic images of our models. 
Fig.~\ref{fig:sed_im} shows the central region of model \texttt{RHD\_MD1e-8} in steady state for an inclination of 45$^\circ$. 
The three panels from top to bottom are synthetic
    images at wavelengths 1.25, 2.2 and 4.8~$\mu$m ($J$, $K$ and $M$
    bands). The field of view is 2~AU wide and includes the hot optical thin gas and the actual rim (compare zone B and C in Section~\ref{sec:model_hydrostatic}). 
The $J$ and $K$ bands show both a bright inner ring at the $\rm \tau_*=1$ midplane region, while the extended triangular cross-section of the rim is cooler and emits more in $M$ band. 
The small dust halo in front of the rim emits slightly in the $J$ band, see Fig.~\ref{fig:sed_im}, top. 
Fig.~\ref{fig:sed_im_2} shows the three bands combined into a
    color image, with $J$, $K$ and $M$ mapped to the blue, green and
    red channels. The plots shows that overall the $K$ and $M$ emission is strongest (red-yellow colors) while the contribution from $J$ bands remains small.
%
%
%
 
%

\section{Discussion}
\label{sec:disc}

This work represents a first step in constructing self-consistent
models of the inner regions of protoplanetary disks that account for
both dynamical and thermodynamical constraints while remaining
numerically tractable. There are still some important limitations to this
work which we will overview in the following. 

We considered a uniform dust-to-gas mass ratio of
0.01. However, both turbulence as well as dust
settling and radial drift all change the local dust density and
consequently the opacity. In addition, small grains could be quickly
depleted due to the fast growth and settling
\citep{bra08,bir10,zso11,oku12}. Our models show a pressure maximum appearing at the location of the ionization temperature $\rm T_{MRI}$ at which the surface density
increases due to the drop of accretion stress. At that location,
larger dust particles could be concentrated and increase the dust-to-gas mass ratio. Collisions between the larger particles could provide the small grains which could be mixed in the
upper layers. It is possible that such an increase in dust density
could help increase the height of the rim and create a larger
shadowed region. However, due to the fact that this location is at
lower temperatures ($\rm T_{MRI} < T_{ev}$), it is located further
outward in the disk and is unlikely to increase the flux at $\rm 2 \mu
m$. In addition, the rim can also be affected by the photoelectric heating \citep{thi11} and the radiation pressure on the dust \citep{vin14}. A more sophisticated treatment of the dust is needed in the future.
%
%

In this work, we have fixed the gas opacity to a small value of $\rm \kappa_{gas}=10^{-4}$ cm$^2$ g$^{-1}$. For this value, the inner gas disk remains optically thin for a given range of mass accretion rates from $\rm \dot{M}=10^{-9}$ to $\rm \dot{M}=10^{-7}$ solar mass per year. 
One way to change the gas opacity without affecting the rim position would be to change the value of the accretion stress in the ionized region. Assuming an accretion stress of $\rm \alpha_{in}=0.1$ for temperatures above 1000~K would reduce the gas surface density by one order of magnitude and so allow a higher gas opacity of $\rm \kappa_{gas}=10^{-3}$ without affecting the optical depth of the inner gas disk. Finally, we note again that a detailed implementation of gas line radiation transfer and the frequency dependent gas opacity would go far beyond the scope of this work.

\begin{figure*}[ht]
\centering
\includegraphics[width=18.0cm]{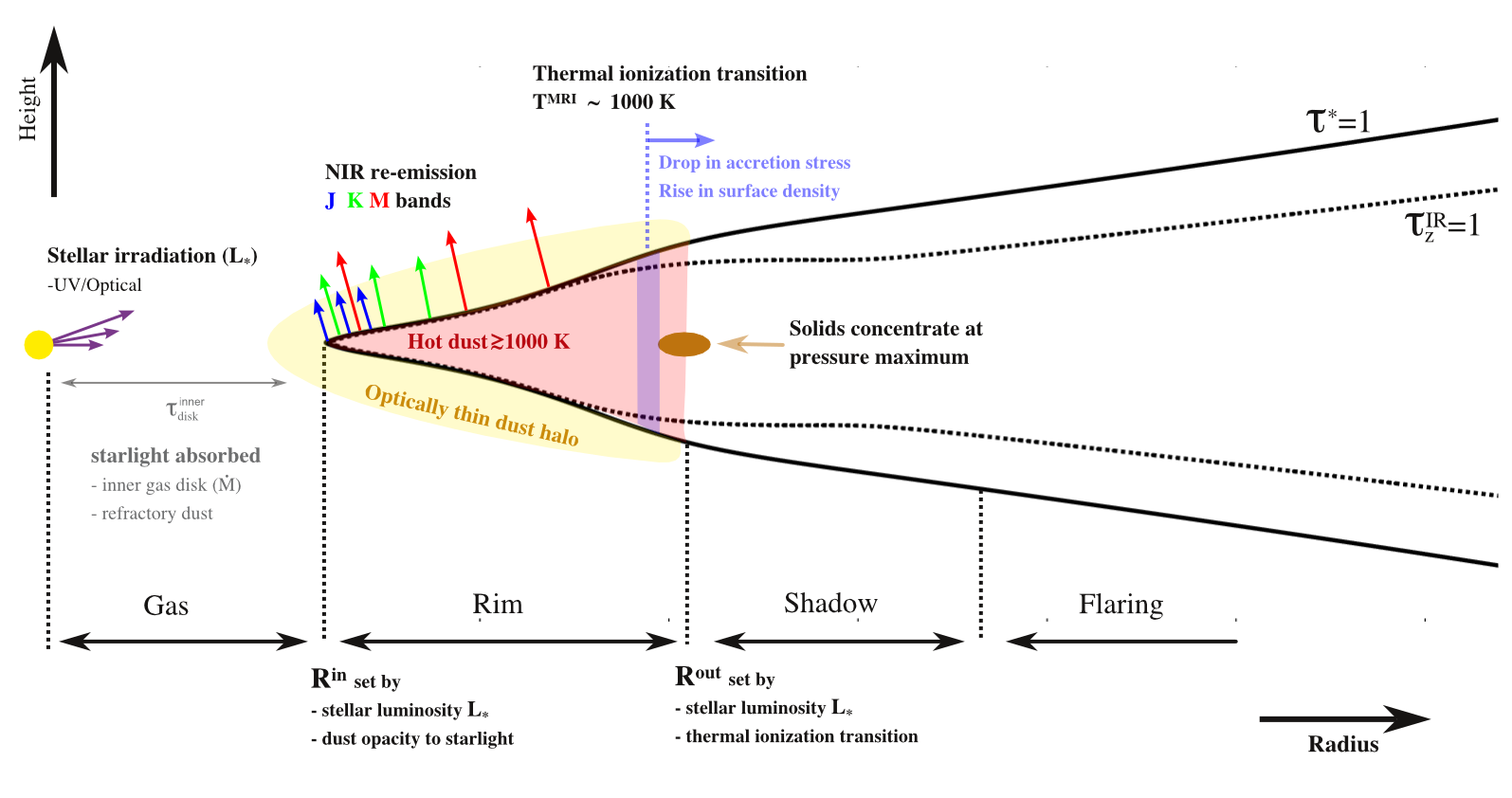}
%
%
\caption{Schematic structure of the inner rim of the disk around a young 
intermediate-mass star, summarizing the results of our radiation hydrodynamical models. This meridional cross-section has the star 
and the system's rotation axis at left. Vertical dotted lines divide the disk into, from left to right, (1) optically-thin gas 
with silicate vapor; (2) the silicate sublimation front, forming the 
inner rim of the optically-thick, dusty material --- the starlight
absorbed at the front heats the rim, making it vertically extended;
(3) the shadow cast by the rim; and (4) the outer disk, whose
slightly upward-curving surface lets it too see the star. The local
pressure maximum just beyond the rim is a location where
planet-forming solids can accumulate at temperatures near 1000~K.}
\label{fig:sketch}
\end{figure*}

\section{Conclusions}
\label{sec:conc}

We have developed the first radiation hydrodynamical models of the
silicate sublimation front in protoplanetary disks around Herbig Ae
stars. The models are axisymmetric and include stellar irradiation,
dust and gas opacity, dust sublimation and condensation. The effects
of turbulence (angular momentum transport and dissipative heating) are
modeled by means of a temperature dependent kinematic viscosity. This
dependence is chosen to capture the onset of magneto-rotational
turbulence due to dust grains' thermionic emission at
temperatures above about 1000~K \citep{des15}. The models 
are inflow-equilibrium solutions with radially-constant mass 
accretion rates. We compute cases with mass flow rates from $\rm \dot{M}=10^{-9}$ to 
$\rm \dot{M}=10^{-7}$ solar masses per year, and stellar luminosities from $\rm L_*=11\, L_\sun$ to $\rm L_*=56\, L_\sun$. 

Using numerical simulations in which we solve the time
dependent radiative-hydrodynamics equations, including viscosity, we
have shown the stability of the rim to axisymmetric modes. In addition, 
the models remain in steady state for thousands of dynamical timescales. 
For $\rm \dot{M} \le 10^{-8}$ solar mass per year, accretion heating has only a modest effect on the flow. The height of the near infrared optical depth unity $\rm
\tau_Z^{NIR}=1$ is at around $\rm Z/R=0.1$ to $\rm Z/R=0.14$,
depending on the model parameters and the mass accretion rate. The
spectral energy distributions of our models, calculated by Monte-Carlo
radiative transfer tools show good agreement with previous radiation
hydrostatic model \citep{mul12,vin14}.

The inner disk can be divided into three structures, summarized in Fig.~\ref{fig:sketch}. The first is an
optically thin halo of hot dust between the sublimation front and the
star. Such hot dust in front of the rim has been proposed in previous models \citep{kam09}, though  its extent and optical depth is lower than predicted by \citet{vin06}. The second
structure is the actual rim front where the dust condenses over the geometrically-thin layer absorbing most of the starlight. This front has a triangular
cross-section in the poloidal plane, with the point lying in the
midplane aimed at the star.  
The triangle cross-section is quite radially extended, stretching several times the gas density scale height. 
The third structure is the material lying outside the sublimation front, which is shadowed from the 
starlight by the optically-thick dust. The radial position of the rim in our models matches well with observational constraints from near-infrared interferometry. These results depend on the parameters as follows:
\begin{enumerate}

\item $\, \, $The radial extent of the inner rim triangular cross-section shape depends mainly on the mass accretion rate. This extent could be several tenth of an AU. Higher mass accretion rates increase this radial extent of the rim. Such an extent is consistent with near-infrared interferometry observations which have clearly shown a smooth, radial extended structure, derived from visibility curves \citep{tan08,ben10}.  Finally, even though the mass accretion rate has an effect on the radial extent it has only a small effect on the spectral energy distribution and the actual radial position of the rim.

\item $\, \, $We find that the accretion heating  only affects the temperature profile in the shadowed region behind the inner rim for the given parameter space we investigated. We observe no substantial increase of the actual inner rim height with increasing accretion rate. For our models, accretion heating becomes important for a mass accretion rate above $\rm \dot{M} \ge 10^{-8}$ solar mass per year. The relative weak dependence on accretion heating can also be explained by the fact that the high accretion stress is expected for temperatures above the ionization temperature of 1000~K and such regions are optically thinner than the regions below 1000~K. 

\item $\, \, $For the given ionization temperature we found that the location of the pressure maxima is at a region with temperatures around 1000~K. Solids are likely to concentrate near this radius under gas drag forces, since the pressure gradients make the gas rotate faster than Keplerian inside the maximum, and slower than Keplerian outside \citep{hag03}. Concentration of pebbles or boulders followed by collisional fragmentation could lead to more abundant sub-micron dust, which would increasing the height where the optical depth falls to unity. At this region, the temperatures are sufficiently high for annealing to form crystalline silicates.

\end{enumerate}

In summary, we have presented the first models which enable to
study the inner dust rim region with radiation hydrodynamical
simulations. The work should be seen as a bridge, connecting previous
highly sophisticated dust Monte Carlo radiation transfer hydrostatic
models of the inner rim and global hydrodynamical simulations of
stratified disks. With a simplified radiation transfer, we capture the
relevant physics and the presented models compare very well with the
observational constraints of ring radius and SED. The NIR emission at 2~$\mu$m is about half as bright relative to the
host star as observed, similar to existing models \citep{mul10}.  
Whether the missing NIR flux can come from a magnetically 
supported atmosphere remains an open question \citep{tur14}. 
Better characterizing the environments near young stars of 
different masses is essential if we are to understand the origins of
the population of close planets.

For the first time we have performed axisymmetric
radiation hydrodynamical simulations of this region and we confirm the
flow's stability over thousands of orbits.  Furthermore, robust
dust abundance and temperature distributions can be obtained on grids
coarse enough that 3-D calculations are now feasible.
%
%
%
\section*{Acknowledgments}

The authors thank Antonella Natta, Wlad Lyra, Rafael Millan-Gabet, Gijs Mulders and Satoshi Okuzumi for useful comments on
the manuscript. We thank Andrea Mignone for supporting and advising us
with the newest PLUTO code. Parallel computations have been performed
on the Genci supercomputer 'curie' at the calculation center of CEA
TGCC and on the zodiac supercomputer at JPL. For this work, Sebastien
Fromang and Mario Flock received funding from the European Research
Council under the European Union's Seventh Framework Programme
(FP7/2007-2013) / ERC Grant agreement nr. 258729. This research was
carried out in part at the Jet Propulsion Laboratory, California
Institute of Technology, under a contract with the National
Aeronautics and Space Administration and with the support of the NASA
Exoplanet Research program via grant 14\-XRP14\_2\-0153. Copyright
2015 California Institute of Technology. Government sponsorship
acknowledged. 

\appendix

\setcounter{table}{0}
\renewcommand{\thetable}{A\arabic{table}}

\section{{\bf A} Dust opacity}
\label{ap:opac}
\begin{figure}
\centering
\includegraphics[width=9cm]{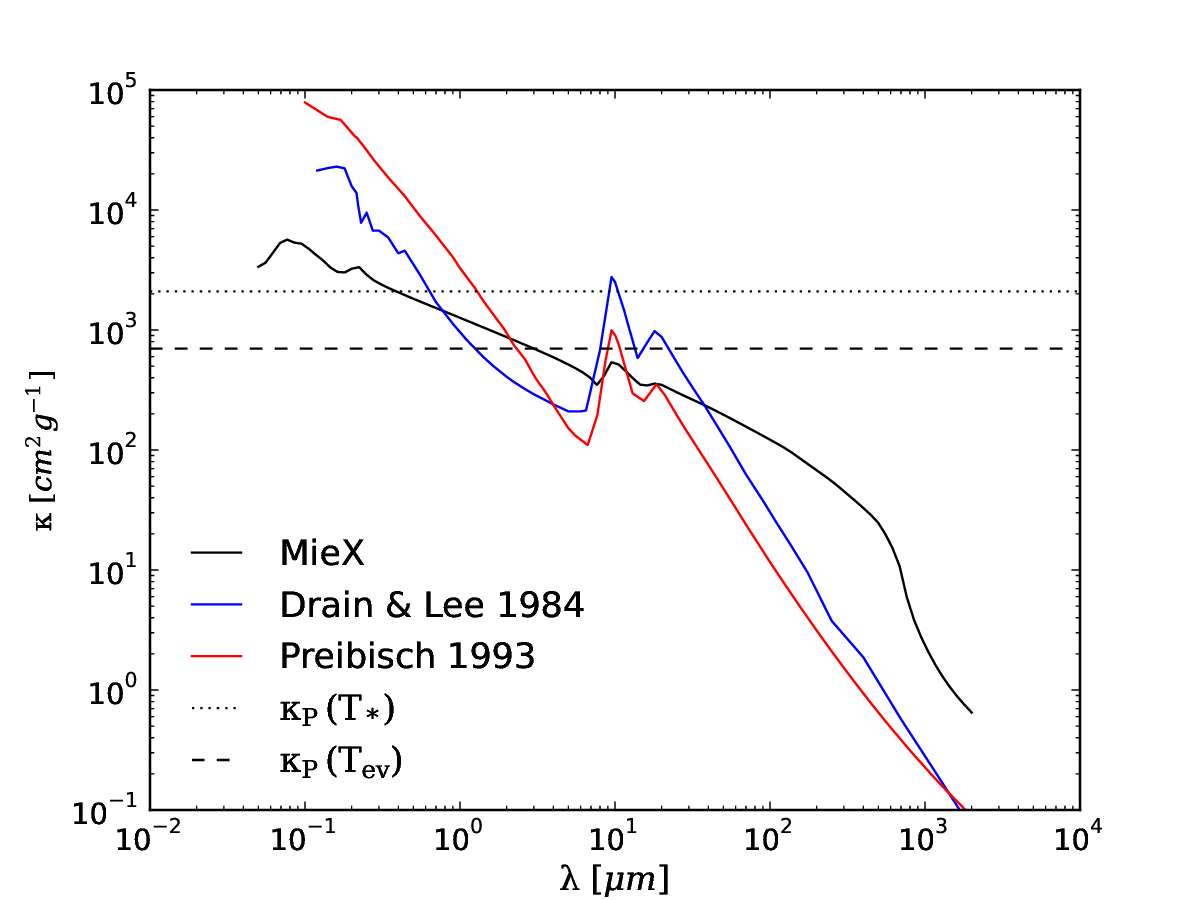}
\caption{Wavelength dependence of the dust opacity. The dust opacity by \citet{dra84} (blue) and \citet{pre93} (red) is plotted for comparison. The two dust opacities used for model \texttt{MDe-8} are overplotted.}
\label{fig:Opac}
\end{figure}

The dust opacities are calculated by Mie theory using the method by \citet{wol04}. To calculate the opacity we assume a range of silicate and carbon particles between ($\rm a_{min}=5 \mu m$ and $\rm a_{max}=100 \mu m$) with a size distribution profile $\rm \sim a^{-3.5}$ and a silicate abundance of 62.5\%. The profile of the dust opacity is plotted in Fig~\ref{fig:Opac}. For comparison we plot also opacity values by \citet{pre93} and by \citet{dra84} which assume slightly smaller dust sizes. To reduce the complexity of the problem we use gray opacities for the simulations. E.g. we define the Planck mean opacity as 
\begin{equation}
\rm \kappa_P(T) = \frac{\int \kappa_\nu B(\nu,T) d\nu}{\int B(\nu,T) d\nu},
\label{eq:planck}
\end{equation}
with the Planck function $\rm B(\nu,T)$. We note that in this work we assume the same evaporation temperature for silicate and carbon grains. Especially refractory carbon grains could survive to higher temperatures. We will address this in a future work.

\section{{\bf B} Test of the dust sublimation function for $\rm T>T_{ev}$}
\label{app:dsf}
\begin{figure}
\centering
\includegraphics[width=9cm]{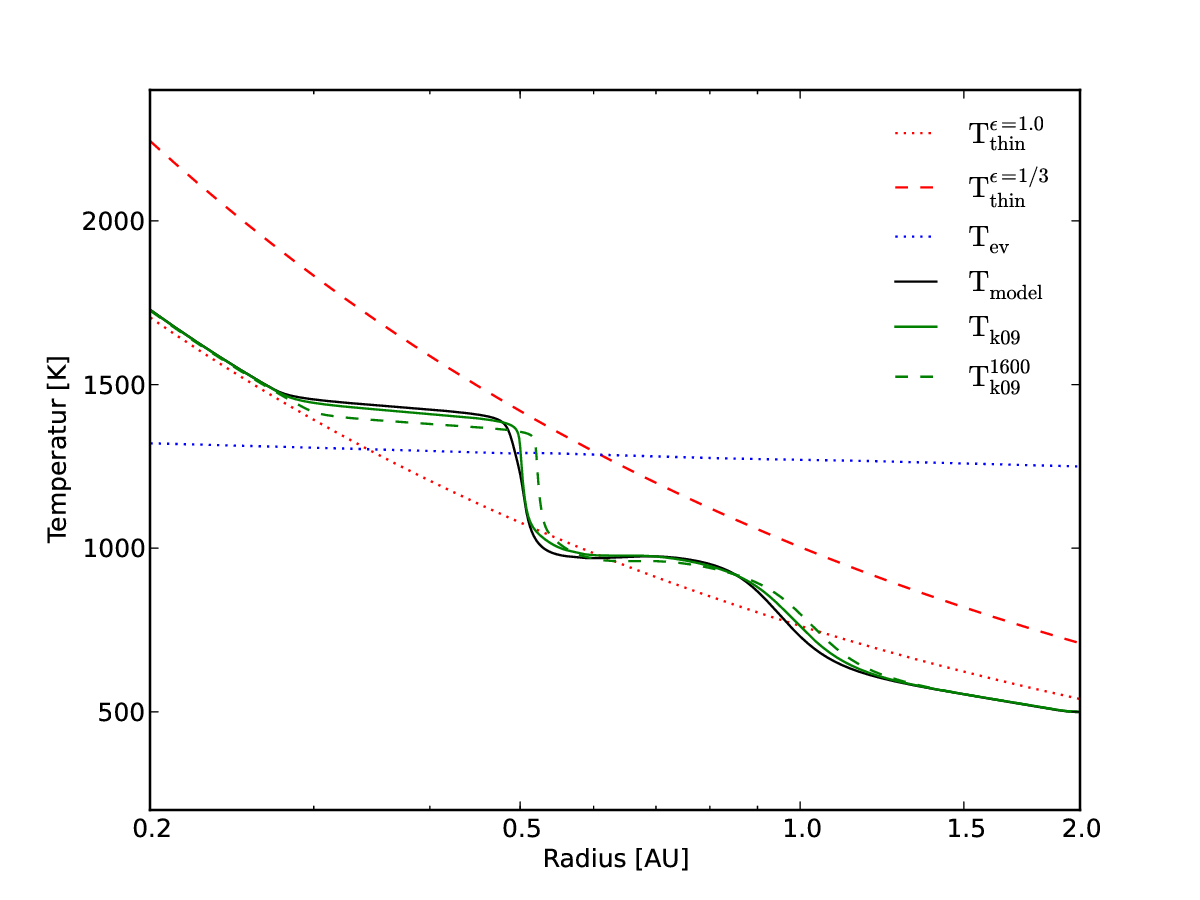}
\caption{Converged radial temperature profiles for different sublimation functions. The red curves correspond to the optical thin temperature of the gas (red dotted line) and the dust (red dashed line). The blue dotted line shows the sublimation temperature of the dust.}
\label{fig:temp_thin_app}
\end{figure}

We perform additional models in which we replace the function for $\rm T>T_{ev}$ in Eq.~(\ref{eq:d2g}) with an exponential function as by \citet{kam09}
\begin{equation}
\rm f^{k09}=exp \left\{ -a_{steep} \left ( \frac{T_{ev}-T}{T} \right )^2 \right\}, 
\label{eq:kama}
\end{equation}
with the steepness factor $\rm a_{steep}$. In the following, we show the influence of the functions in Eq.~(\ref{eq:d2g}) on the converged dust temperature profile. The first function for $\rm T>T_{ev}$ causes a fast sublimation of the dust for temperatures higher than the sublimation temperature. This function, has only a minor influence on the final temperature profiles, compared to different functions with different steepness, see Fig.~\ref{fig:temp_thin_app}. Independent of our choice of evaporation function, the models converge always to a solution with an optical thin dust halo in front of the rim (compare the region with higher temperature in front of the rim). The function for $\rm T<T_{ev}$ in Eq.~(\ref{eq:d2g}) smooths out the high opacity jump due to the dust condensation and so allow us to resolve the irradiation absorption at the rim front independent of the grid resolution and optical thickness. An example of a calculation without using this function is presented in a later section. We also note that the value of $\rm f_{\Delta \tau}$ effects only the steepness of the temperature drop between the $\rm \tau_*=1$ and $\rm \tau_*=100$ position.
In Fig.~\ref{fig:temp_thin_app} we perform additional radiation hydrostatic versions of model \texttt{S100}, following Eq.~(\ref{eq:kama}) and using two different steepness factors of $\rm a_{steep}=400$ and $\rm a_{steep}=1600$. The plot shows that the functions have only a minor effect on the location of the rim. 
%
%
In addition, independent of the steepness factor we naturally obtain an optically thin dust halo in front of the rim which can be seen
at the region with high temperature $\rm T>T_{thin}^{\epsilon=1}$. 
\section{{\bf C} Test of the dust sublimation function for $\rm T<T_{ev}$}
\label{app:subfunc}
\begin{figure}
\centering
\includegraphics[width=8.5cm]{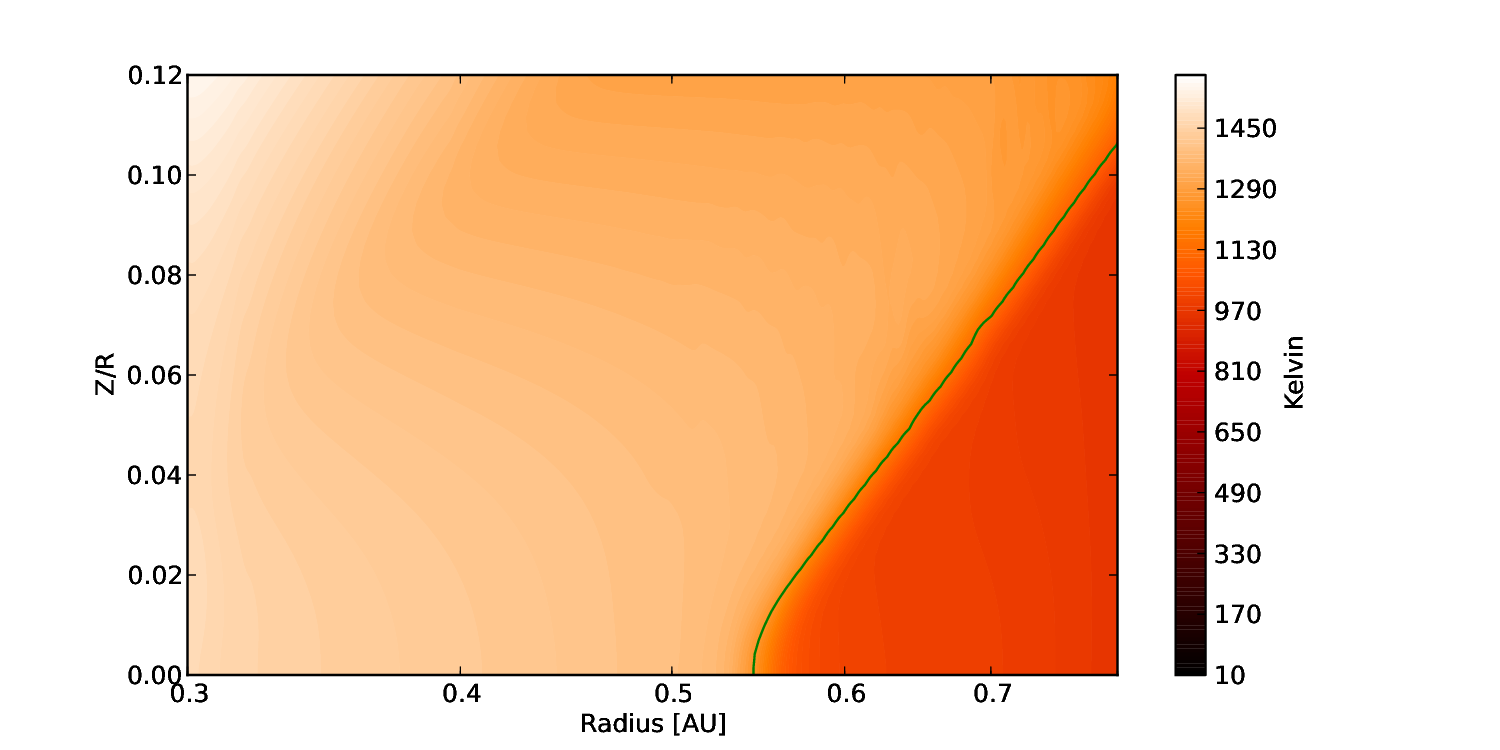}
\includegraphics[width=8.5cm]{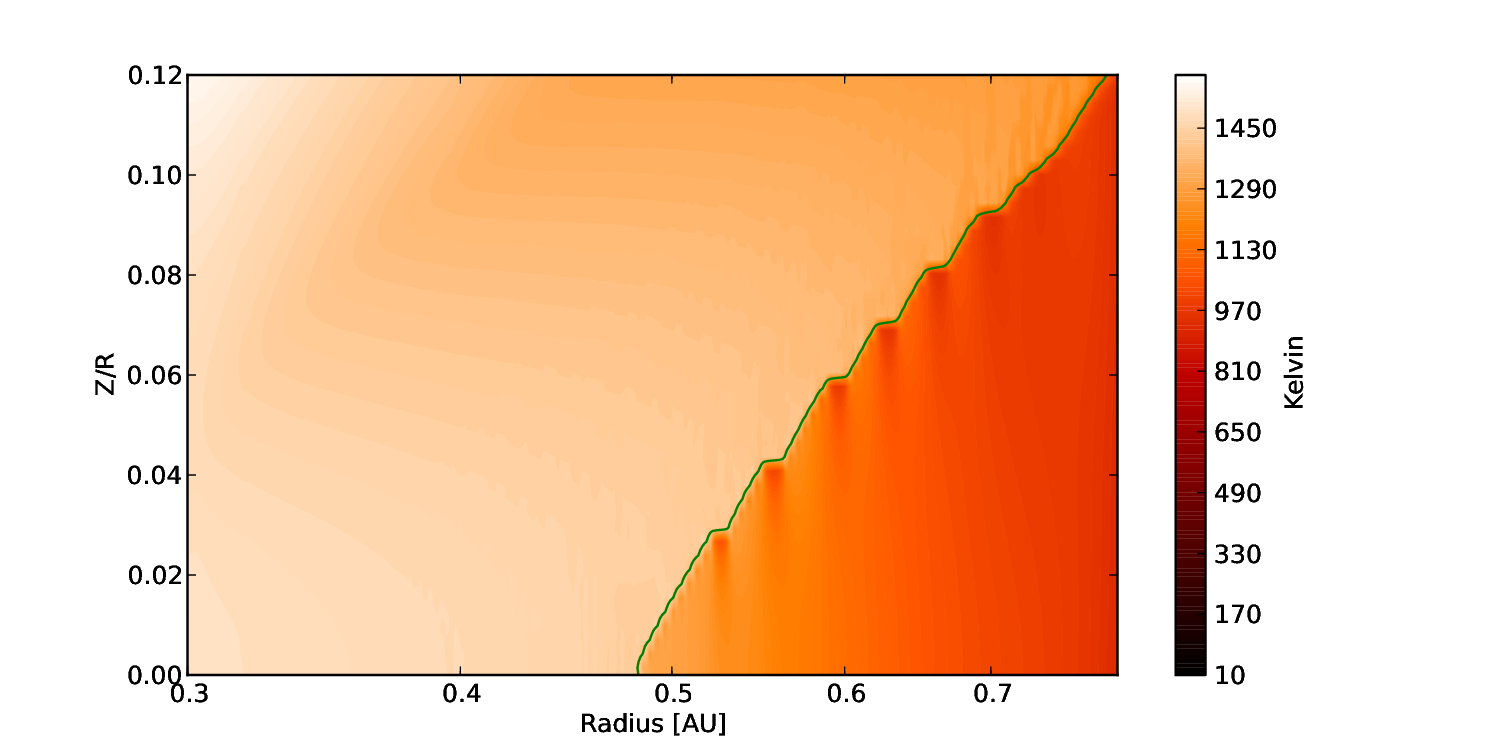}
%
\caption{Comparison of 2D temperature profile in the R-Z/R plane, after 6 inner orbits for model \texttt{RHD\_S100} (left) and model \texttt{RHD\_S100$^{MOD}$} (right). The green lines indicate the optical depth unity for the irradiation.}
\label{fig:templog}
\end{figure}

In this test, we want to show the importance of the dust sublimation function Eq.~(\ref{eq:d2g}) for $\rm T<T_{ev}$. We performed the model \texttt{RHD\_S100} which uses as initial conditions the results of model \texttt{S100}. We also performed a model without the second function in Eq.~(\ref{eq:d2g}), here called model \texttt{RHD\_S100$^{MOD}$}.
Fig. \ref{fig:templog} compares the two snapshots of the 2D temperature profile of model \texttt{RHD\_S100} and model \texttt{RHD\_S100$^{MOD}$} after 200 inner orbits. 
In model \texttt{RHD\_S100$^{MOD}$}, the irradiation heating is absorbed in one cell which causes jumps in the temperature profile. We note that this effect becomes even stronger for simulations with higher surface density and so higher optical depth. With this test we want to underline the importance to resolve the irradiation absorption and therefor the use of Eq.~(\ref{eq:d2g}).
%


\section{{\bf D} Frequency dependent irradiation and temperature dependent dust opacity}
\label{ap:freq}
\begin{figure}
\centering
\includegraphics[width=8.5cm]{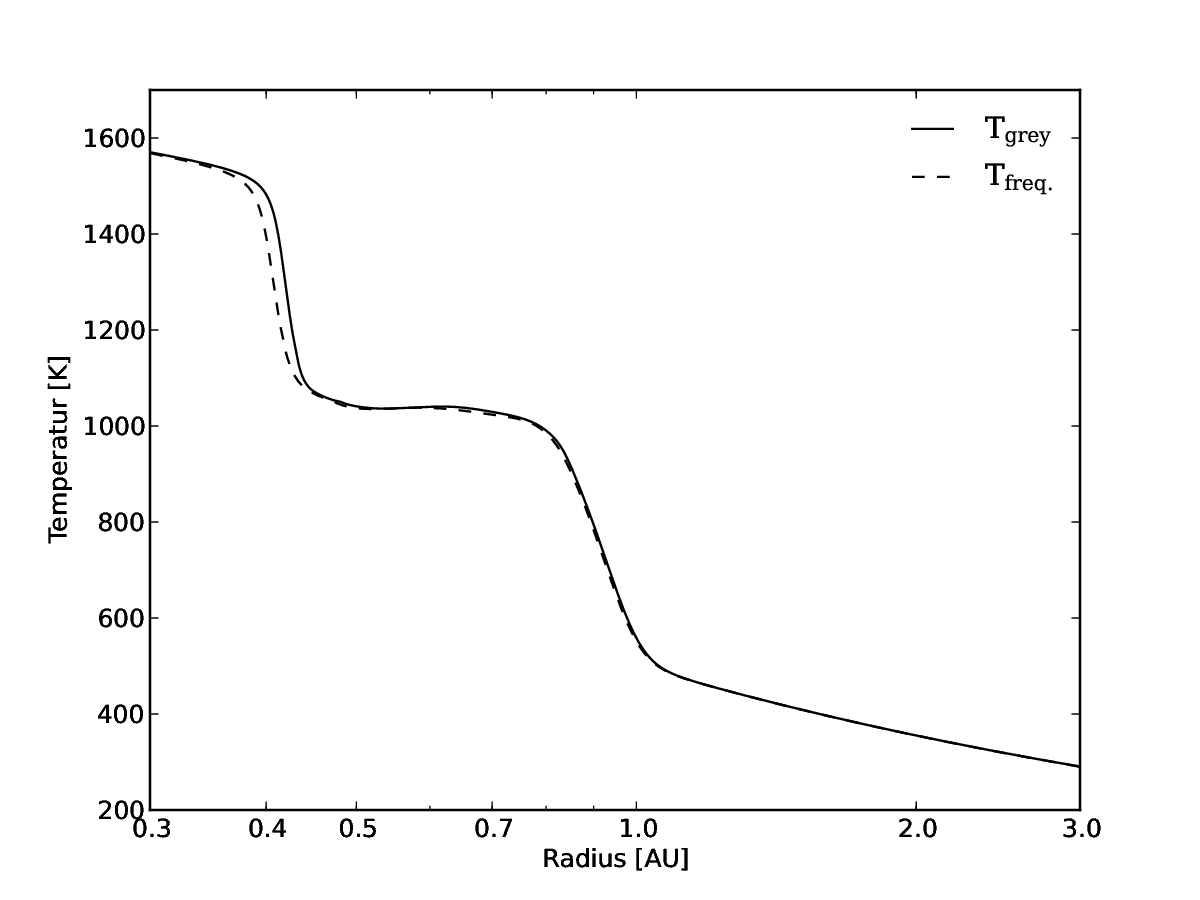}
\includegraphics[width=8.5cm]{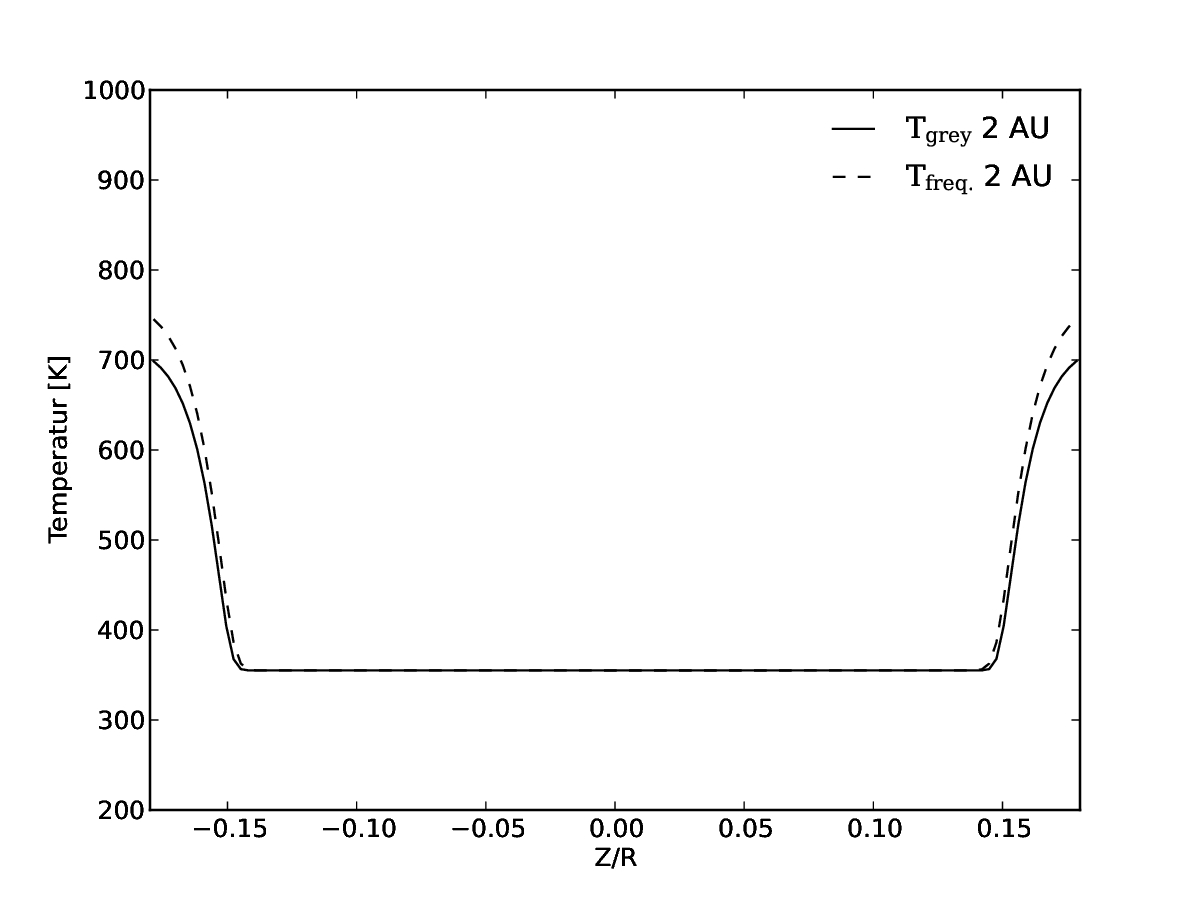}
%
%
\caption{Temperature profiles of model \texttt{MD1e-8} using full frequency dependent irradiation and temperature dependent dust opacity (dashed line) compared to the model using a constant two opacity model (solid line) over radius (left) and height (right).}
\label{fig:2dtemp_freq}
\end{figure}

In this subsection we compare our simplified constant two opacity
model, described in Section~\ref{sec:method} with a more complex model, using
frequency dependent irradiation and Planck mean dust opacity which
depends on temperature $\rm \kappa_P(T)$. For further details, e.g. on the calculation of the
frequency dependent irradiation we refer to our previous work \citep{flo13}.
For this additional model we use 60 frequency bins to sample the opacity, presented in
Fig.~\ref{fig:Opac}. In addition, the Planck and Rosseland opacity is
calculated in each cell for the given dust amount and dust
temperature. The resulting midplane and vertical temperature profiles for the two models are
presented in Fig.~\ref{fig:2dtemp_freq}. There is no significant
difference visible between those two models, both in terms of
temperature as well as structure of the rim. The fixed opacity model reproduces very well the structure and temperature
profile of the more complex model using frequency dependent
irradiation. We note that in the fixed opacity model, we
overestimate the local Planck opacity in a large area in the disk by a
factor between 1 and 2.   

\section{{\bf E} Modifications of the dynamical RHD setup}
\label{ap:rmhd}
The 2D radiation hydrodynamical simulations we present in Section~\ref{sec:2D_sim} were performed using a second order in 
space and time numerical configuration of the PLUTO code
\citep{mig12a}. We solve the same RHD equations as in \citet{flo13},
with two modifications: the magnetic field is set to zero and we
include in the momentum and total energy equations the effect of a
finite kinematic viscosity. Its amplitude is calculated as described
in Section~\ref{sec:surf_dens}. We used the Harten-Lax-Van Leer (HLL)
Riemann solver with a Courant number of 0.3 to increase the 
numerical stability. 

The domain size is set to $\rm R_{in} - R_{out}  = 0.3 -3.0$~AU and $\rm
\Delta \theta = 0.36$ radian and we used $\rm 1024 \times 128$ grid
cells, respectively in the radial and meridional directions. The
initial conditions are provided by the radiation hydrostatic models
\texttt{MDe-9} and \texttt{MDe-8} and are interpolated on the
computational grid using the built in interpolation routine in PLUTO
that allows to change the domain size and grid resolution. 

Three modifications were necessary to increase the numerical stability
of the simulations:
\begin{enumerate}[topsep=0pt,leftmargin=10pt,itemsep=0pt]
\item[$\bullet$] $\,$ A buffer zone extending from 0.3 to 0.32~AU and
  2.5~AU to 3.0~AU in radius, where we reestablish the surface density
  on a timescale shorter than the dynamical timescale. This prevents
  the loss of material due to the pure outflow radial boundary
  condition. 
\item[$\bullet$] $\,$ A time adaptation routine for the dust
  sublimation which smooths out strong fluctuations in the local dust
  amount for each time integration step. As a result, changes of the
  dust density are limited to $10\%$ per timesteps. This prevents sudden
  changes in the opacity that could lead to problems for the radiative
  solver to converge.  
\item[$\bullet$] $\,$ A modified gravitational potential in the radial
  inner and vertical upper layers. This was done by modifying the
  gravitational potential according to
\begin{equation}
\rm \Phi = \frac{G M_*}{r- 0.175 ( 0.4 AU - r) (\theta -\pi/2)^2}\, \, for\, r < 0.4 AU\\
\end{equation}
\begin{equation}
\rm \Phi = \frac{G M_*}{r}\, \, elsewhere.
\end{equation}
Its effect is to reduce the density contrast from $13$ to $9$ orders
of magnitude. We found that this small change in the uppermost
layers substantially increases the stability of the dynamical
calculations. 
\end{enumerate}

\section{{\bf F} Resolution test}
\label{app:res}

\begin{figure}
\centering
\includegraphics[width=8.5cm]{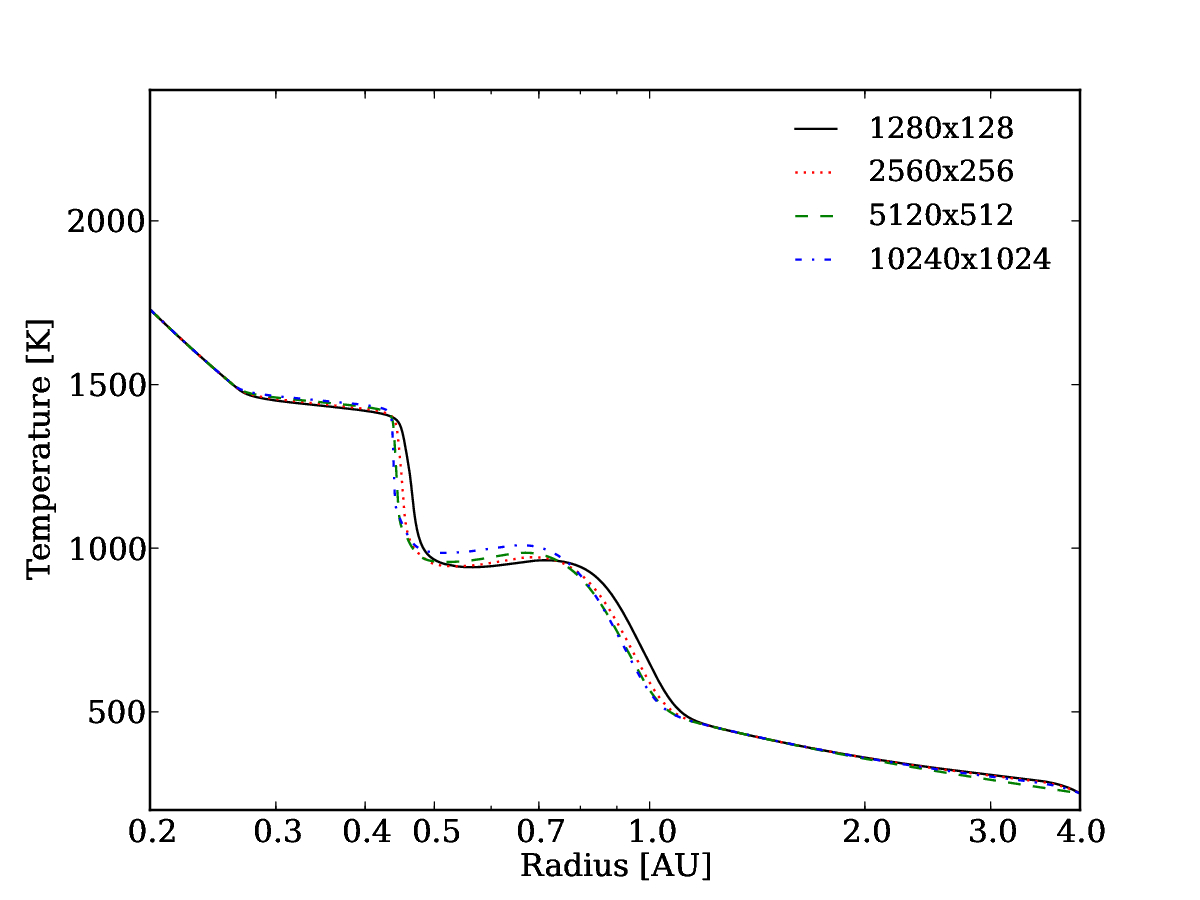}
\includegraphics[width=8.5cm]{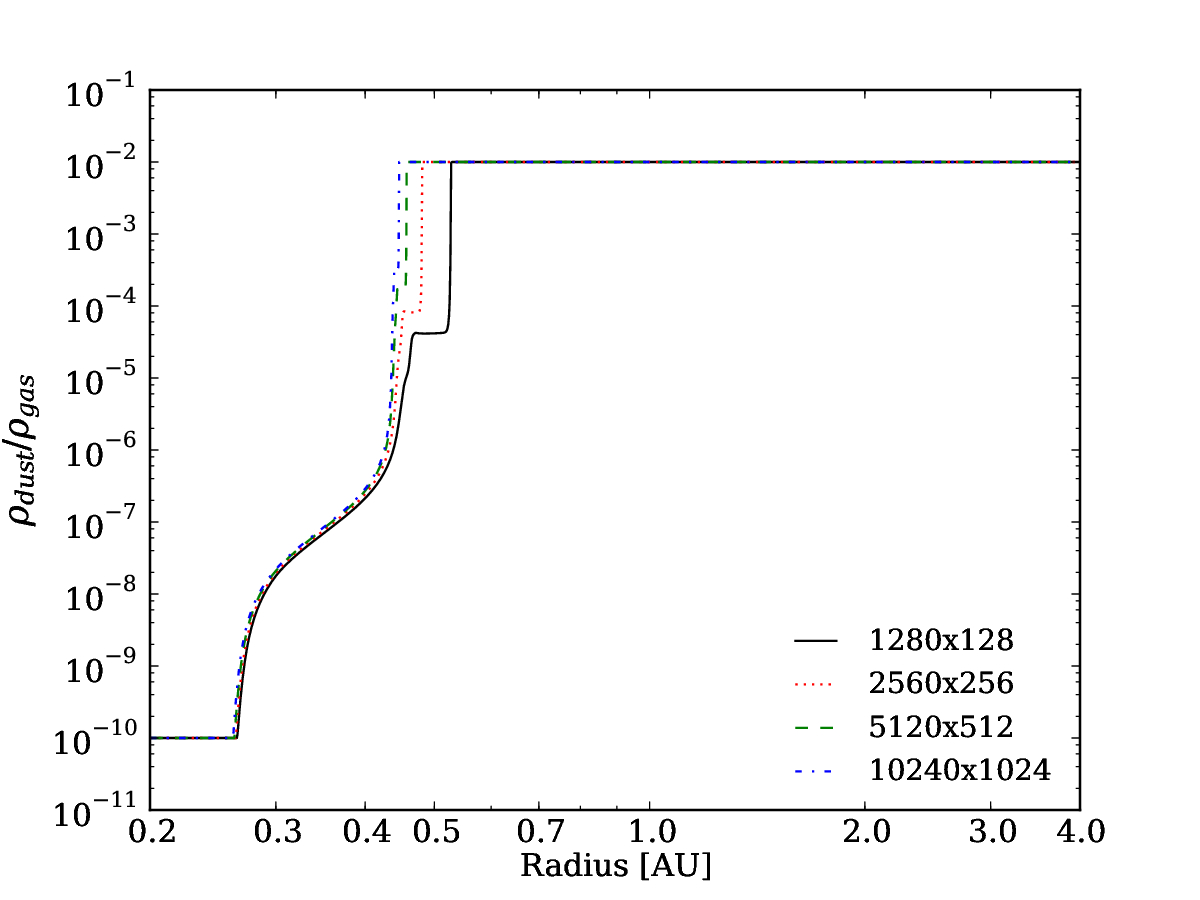}
\includegraphics[width=18.0cm]{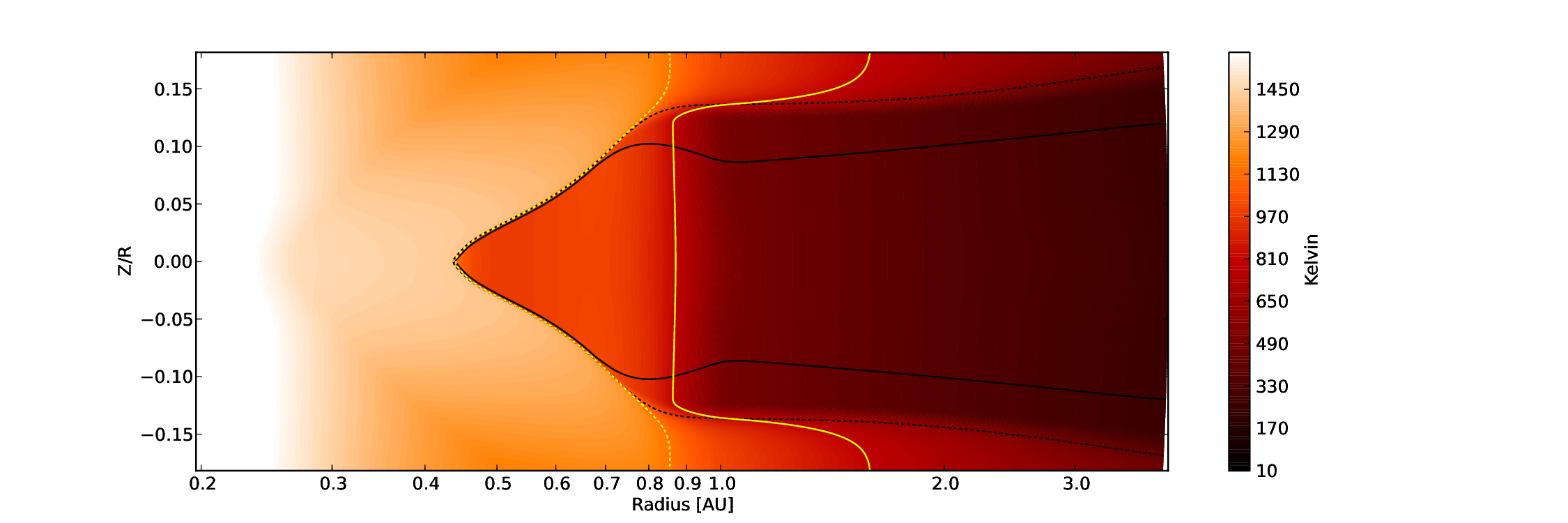}
%
%
\caption{Radial midplane temperature (top left) and radial dust-to-gas mass ratio (top right) for different resolutions. Bottom: 2D temperature contour plot for the highest resolution model (10240 x 1024 cells).}
\label{fig:res}
\end{figure}

To check whether the rim structure is robust against changes in the
  spatial resolution, we perform a convergence study on model \texttt{S100}.
  The three new hydrostatic solutions have resolutions 2, 4 and 8
  times finer than the model described in Section~\ref{sec:model_hydrostatic}.  Fig.~\ref{fig:res} shows that
  the higher the resolution, the thinner the layer near unit starlight
  optical depth where the dust abundance is limited by Eq.~\ref{eq:d2g}, and the
  steeper the temperature gradient.  However, over this eightfold
  range in spatial resolution, the temperatures away from the front
  vary by at most a few percent.  We conclude that despite the
  smoothing applied to the dust abundance near the rim, our
  calculations recover the correct overall structure.

\begin{table}
\begin{tabular}{ll}
\hline
$\rm \mu_g = 2.353$ & Mean molecular weight\\
$\rm k_B= 1.3806 \times 10^{-16}$ erg K$^{-1}$ & Boltzmann constant\\
$\rm u = 1.6605 \times 10^{-24} $ g & Atomic mass unit\\
$\rm G = 6.6726 \times 10^{-8} $ cm$^{3}$g$^{-1}$s$^{-2}$ & Gravitational constant\\
$\rm a_R= 7.5657 \times 10^{-15}$ erg cm$^{-3}$ K$^{-4}$ & Radiation constant\\
$\rm \sigma_b=5.6704 \times 10^{-5}$ erg cm$^{-2}$s$^{-1}$K$^{-4}$ &
Stefan-Boltzmann const. \\
$\rm c =2.99792 \times 10^{8}$ cm s$^{-1}$ & Speed of light\\
\hline
\end{tabular}
\caption{Physical constants used in the work.}
\label{tab:constants}
\end{table}

\bibliographystyle{apj}
\bibliography{RIM}

\end{document}